\documentclass[%
 reprint,
superscriptaddress,
nofootinbib,
 amsmath,amssymb,
 aps,
twocolumn,
]{revtex4-2}

\usepackage{graphicx}
\usepackage{dcolumn}
\usepackage{bm}
\usepackage{float}
\usepackage[table,x11names]{xcolor}
\usepackage{tabularx}

\newcommand{\kcb}[1]{\textcolor{black}{#1}}

\usepackage{esint}
\usepackage{siunitx}
\usepackage{multirow}
\usepackage{booktabs}
\usepackage{makecell}
\usepackage[normalem]{ulem}
\usepackage{soul}
\usepackage[normalem]{ulem}
\setstcolor{red}
\usepackage{cancel}

\usepackage{tikz}
\usetikzlibrary{tikzmark}

\definecolor{rose}{rgb}{1,0.8,0.8}

\begin{document}

\preprint{APS/123-QED}

\title{Piezo-optomechanical signal transduction using Lamb wave supermodes in a suspended Gallium Arsenide photonic integrated circuits platform}
\author{Ankur Khurana}
\email{ankur.khurana@bristol.ac.uk}
\address{Center for Doctoral Training in Quantum Engineering, H.H. Wills Physics Laboratory, Tyndall Avenue, University of Bristol, BS8 1FD, United Kingdom}
\address{Quantum Engineering Technology Labs and Department of Electrical and Electronic Engineering, University of Bristol,
Woodland Road, Bristol BS8 1UB, United Kingdom}

\author{Pisu Jiang}
\address{Quantum Engineering Technology Labs and Department of Electrical and Electronic Engineering, University of Bristol,
Woodland Road, Bristol BS8 1UB, United Kingdom}

\author{Krishna C. Balram}
\email{krishna.coimbatorebalram@bristol.ac.uk}
\address{Quantum Engineering Technology Labs and Department of Electrical and Electronic Engineering, University of Bristol,
Woodland Road, Bristol BS8 1UB, United Kingdom}

\date{\today}

\begin{abstract}

Piezoelectric optomechanical platforms present one of the most promising routes towards efficient transduction of signals from the microwave to the optical frequency domains. New device architectures need to be developed in order to achieve the stringent requirements for building efficient quantum transducers. In this work, we utilize the mechanical supermode principle to improve the overall microwave to optical transduction efficiency, by fabricating Lamb wave resonators that are hybridized with the mechanical breathing modes of a rib waveguide in a suspended gallium arsenide (GaAs) photonic integrated circuits (PIC) platform. Combining the strong elasto-optic interactions available in GaAs with the increased phonon injection efficiency enabled by this architecture, we demonstrate signal transduction up to 7 GHz, and an increase in transduction efficiency by $\approx$ 25$\times$ for the hybridized mode ($f_m\approx$ 2 GHz), using this approach. We also outline routes for improving device performance to enable quantum transduction within this platform. 


\end{abstract}

\maketitle


\section{\label{sec:level1}INTRODUCTION}

Engineering hybrid quantum systems \cite{kurizki2015quantum}, that combine a ``best of all worlds" approach, brings with it the necessity of engineering efficient quantum transducers that can translate the information back and forth between the different physical (frequency) domains. A prototypical example of such a hybrid system is a distributed quantum network built around superconducting qubits, where the distant nodes are linked by low-loss optical fibers, with the superconducting qubits being used for computation and telecom band optical photons for communicating the quantum information between distant nodes. Such a system requires efficient microwave to optical signal transducers
and has attracted a lot of recent research interest \cite{han2021microwave}, primarily driven by the spectacular progress in scaling superconducting qubit based quantum processors \cite{arute2019quantum,gong2021quantum}. In addition to quantum transduction, these devices are also of interest to enable optical detection of weak microwave signals \cite{bagci2014optical} arising in radio astronomy and nuclear magnetic resonance \cite{takeda2018electro}.

The key challenge with building efficient microwave to optical signal transducers lies in overcoming the massive disparity in frequencies (GHz - 100s THz) and associated wavelengths (\SI{}{\centi\meter} - \SI{}{\micro\meter}), which reduces the interaction strength between the fields considerably. By converting the microwave signal to an acoustic wave which has $\approx$ \SI{}{\micro\meter} wavelength at GHz frequencies, piezoelectric optomechanical transducers overcome this wavelength mismatch problem and present one of the most promising routes towards building microwave to optical signal transducers \cite{mirhosseini2020superconducting, balram2022piezoelectric}, by engineering strong acousto-optic interactions in nanoscale optomechanical cavities \cite{aspelmeyer2014cavity}. Such devices have been explored across a wide range of material platforms ranging from monolithic implementations in aluminum nitride (AlN) \cite{bochmann2013nanomechanical, vainsencher2016bi,  li2015nanophotonic, ghosh2016laterally, han2020cavity} , gallium arsenide (GaAs) \cite{balram2016coherent, forsch2020microwave}, gallium phosphide (GaP) \cite{honl2021microwave, stockill2021ultra} and lithium niobate (LN) \cite{shao2019microwave, jiang2020efficient} to hybrid material platforms, AlN-on-silicon (Si) \cite{mirhosseini2020superconducting} and Si-on-LN \cite{marinkovic2021hybrid}. 

In a piezoelectric optomechanical platform, the microwave to optical signal transduction is carried out in two steps. The microwave signal is first converted to an acoustic wave and this acoustic wave then either directly or indirectly excites the mechanical mode of an optomechanical cavity. An optical pump field circulating in the cavity is modulated by this acousto-optic interaction and produces an optical sideband, which is the desired signal of interest. The overall transduction efficiency (${\eta_\text{peak}}$) \cite{wu2020microwave} in such piezo-optomechanical devices scales as ${\eta_\text{peak}}\!\propto\!\eta_{\scriptscriptstyle\text{PIE}}C_\text{om}$, where $\eta_{\scriptscriptstyle\text{PIE}}$ represents the phonon injection efficiency, defined as the fraction of the input microwave energy that is transmitted into the desired mechanical mode of the optomechanical cavity and $C_\text{om} \!\propto\! g_{0}^2$ represents the optomechanical cooperativity, which is dependent on the optomechanical coupling rate $g_{0}$\footnotemark[1]. To achieve high $\eta_\text{peak}$, it is therefore critical to design piezoelectric optomechanical platforms that simultaneously enhance $\eta_{\scriptscriptstyle\text{PIE}}$ and $C_\text{om}$, and this balance is not easy to achieve \cite{balram2022piezoelectric}. In particular, it is challenging to achieve high $\eta_{\scriptscriptstyle\text{PIE}}$ with high $g_{0}$ optomechanical cavities like 1D photonic crystals \cite{balram2014moving} that exploit strong elasto-optic interactions. This occurs \cite{balram2022piezoelectric} because the piezoelectric transducer needs to be simultaneously impedance matched to \SI{50}{\ohm} and mode-matched to $\approx$ \SI{}{\micro\meter}-scale dimensions. It was recently proposed \cite{wu2020microwave} that mechanical supermodes can in principle overcome this challenge. A mechanical supermode, engineered by hybridizing a standard RF-MEMS resonator with an optomechanical cavity with high $g_{0}$, provides a best of both worlds approach by enabling impedance matching through the RF-MEMS front-end and a sufficiently strong optomechanical interaction in the high $g_0$ back-end. The mode hybridization, if properly engineered, allows one to efficiently transfer energy between the two sub-systems and significantly improve the overall $\eta_{\scriptscriptstyle\text{PIE}}$ \cite{wu2020microwave}.

In this work, we design and fabricate Lamb wave resonators in a suspended GaAs photonic integrated circuits platform \cite{jiang2020suspended} with a view towards improving $\eta_{\scriptscriptstyle\text{PIE}}$ for microwave to optical signal transduction, by hybridizing the Lamb wave resonance with the mechanical breathing mode of a rib waveguide. The GaAs platform supports strong acousto-optic interactions with $g_0>$ \SI{1}{\mega\hertz} in 1D nanobeam photonic crystal cavities \cite{balram2014moving}. On the other hand, its weak piezoelectric coefficient ($k^2_\text{eff}$) has limited the achievable phonon injection efficiency, with  $\eta_{\scriptscriptstyle\text{PIE}}\ll$ 1,  which has resulted in overall photon transduction efficiencies of $\approx 10^{-10}$ \cite{balram2022piezoelectric} in experiments. Improving $\eta_{\scriptscriptstyle\text{PIE}}$ is therefore critical to achieving quantum transduction in this platform and the mechanical supermode approach provides a natural route towards higher $\eta_{\scriptscriptstyle\text{PIE}}$ and we show in this work an $\approx$ 25$\times$ improvement in a proof-of-principle experiment, with a potential for higher gains with better device engineering. Lamb wave resonators also enable access to higher mechanical frequencies on account of the higher acoustic velocity of Lamb wave modes, compared to surface acoustic waves (SAW) and we demonstrate transduction of signals up to 7 GHz using this approach. 
\footnotetext[1]{$g_{0}$ has the units of angular frequency (rad.Hz), but $g_0/2\pi$ is more often used which is specified in Hz. In this article, $g_0$ also represents $g_0/2\pi$ wherever the units Hz are used.}

\section{Enhancing phonon injection with Lamb wave resonators}
\begin{figure}[t]
\includegraphics[scale=0.54]{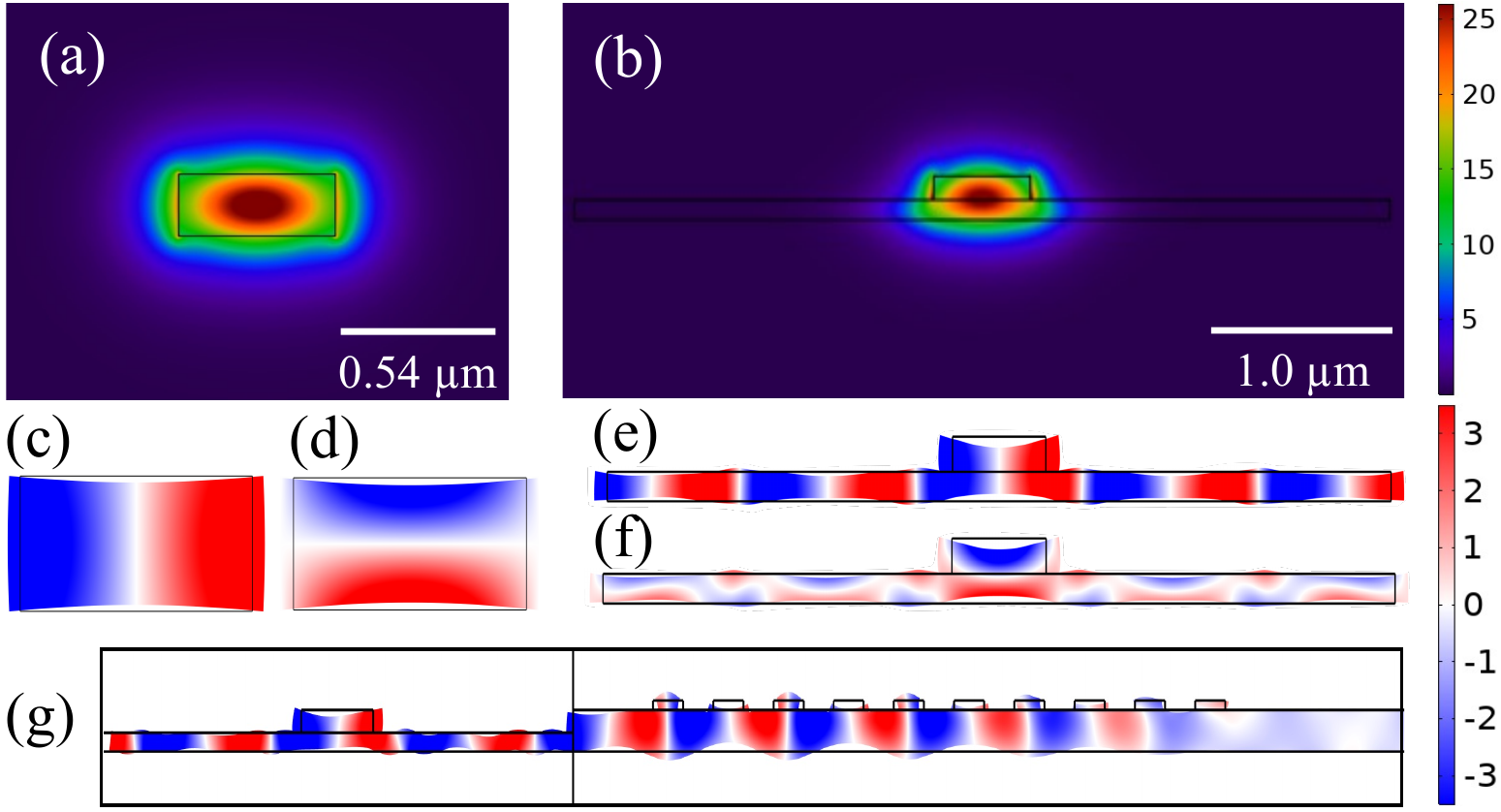}
\caption{Optical modes (TE) of a strip waveguide (a) and a rib waveguide (b) of width \SI{540}{\nano\meter} and thickness \SI{220}{\nano\meter} and total rib width of \SI{3.54}{\micro\meter}, (c),(d) X and Y displacements, respectively, of the shear horizontal (SH) breathing mode (4.2 GHz) for the strip waveguide with $g_0 \approx$ 200 kHz, (e),(f) X and Y displacements, respectively, of the SH breathing mode ($\approx$ 4.5 GHz) for the rib waveguide, (g) X displacement of the SH breathing mode of the rib waveguide being excited by the Lamb wave resonator with an IDT period of  \SI{0.9}{\micro\meter} and 5 finger pairs.} \label{fig:eigenmodes} 
\end{figure}

While the mechanical supermode principle \cite{wu2020microwave} was originally proposed in the context of increasing $\eta_{\scriptscriptstyle\text{PIE}}$ for the breathing mode of a 1D nanobeam optomechanical crystals ($g_0\approx$ \SI{1}{\mega\hertz} in GaAs), in this work we primarily focus on the breathing mode (shear horizontal) of rib waveguides. As shown in Fig.\ref{fig:eigenmodes}(c,d), the shear horizontal mode of a strip waveguide has the same symmetry as the breathing mode a nanobeam cavity, and therefore shows a similarly high $g_0$ ($\approx$ 200 kHz), for a perpendicular length of \SI{60}{\micro\meter} to the 2D cross-section, in overlap with the fundamental transverse electric (TE) mode of the waveguide, shown in Fig.\ref{fig:eigenmodes}(a). The strong acousto-optic coupling of these mechanical waveguide modes has also been exploited for Brillouin scattering experiments in suspended silicon waveguides \cite{rakich2012giant, kittlaus2016large, otterstrom2018silicon}. Moving from a strip waveguide to a rib waveguide (Fig.\ref{fig:eigenmodes}(e,f)) lowers the acousto-optic mode overlap and $g_0$ (with the rib waveguide TE mode shown in Fig.\ref{fig:eigenmodes}(b)), but allows us to excite the mode laterally from a Lamb wave resonator. As Fig.\ref{fig:eigenmodes}(g) shows, by building a Lamb wave (contour mode) resonator \cite{piazza2006piezoelectric} on the other side of the rib and choosing the transducer period to match the rib waveguide breathing mode frequency, $\eta_{\scriptscriptstyle\text{PIE}}$ can be significantly increased due to mode hybridisation. By wrapping the rib waveguide into a microring resonator, the optomechanical interaction can be resonantly enhanced. 

The Lamb wave resonator is designed by building a traditional interdigitated transducer (IDT) on a suspended membrane. A Lamb wave resonator \cite{piazza2006piezoelectric} is a hybrid of traditional surface acoustic wave (SAW) and bulk acoustic wave (BAW) resonators, built by combining the benefits of both. Lamb wave resonators combine the lithographic frequency definition of SAW devices, which enable easy multi-frequency operation, with the higher mechanical quality factors ($Q_m$) at GHz frequencies usually attained in BAW and film bulk acoustic wave resonators (FBAR) devices, enabled by suspension. In contrast to using SAWs for exciting optomechanical platforms \cite{balram2016coherent}, Lamb waves have two major advantages: Lamb waves are better mode matched to the opto-mechanical modes of interest, which increases the $\eta_{\scriptscriptstyle\text{PIE}}$ and overall $\eta_{peak}$ and the higher velocity of Lamb waves compared to SAW waves (2-3$\times$ higher) allows access to higher frequency mechanical modes. The key challenge with fabricating Lamb wave resonators, especially in a weak piezoelectric like GaAs, is the need to suspend a relatively large (size $\approx \SI{1000}{\micro\meter}^2$) membrane and maintaining mechanical integrity across the whole suspended PIC platform.

\begin{figure*}
\includegraphics[scale=1.0]{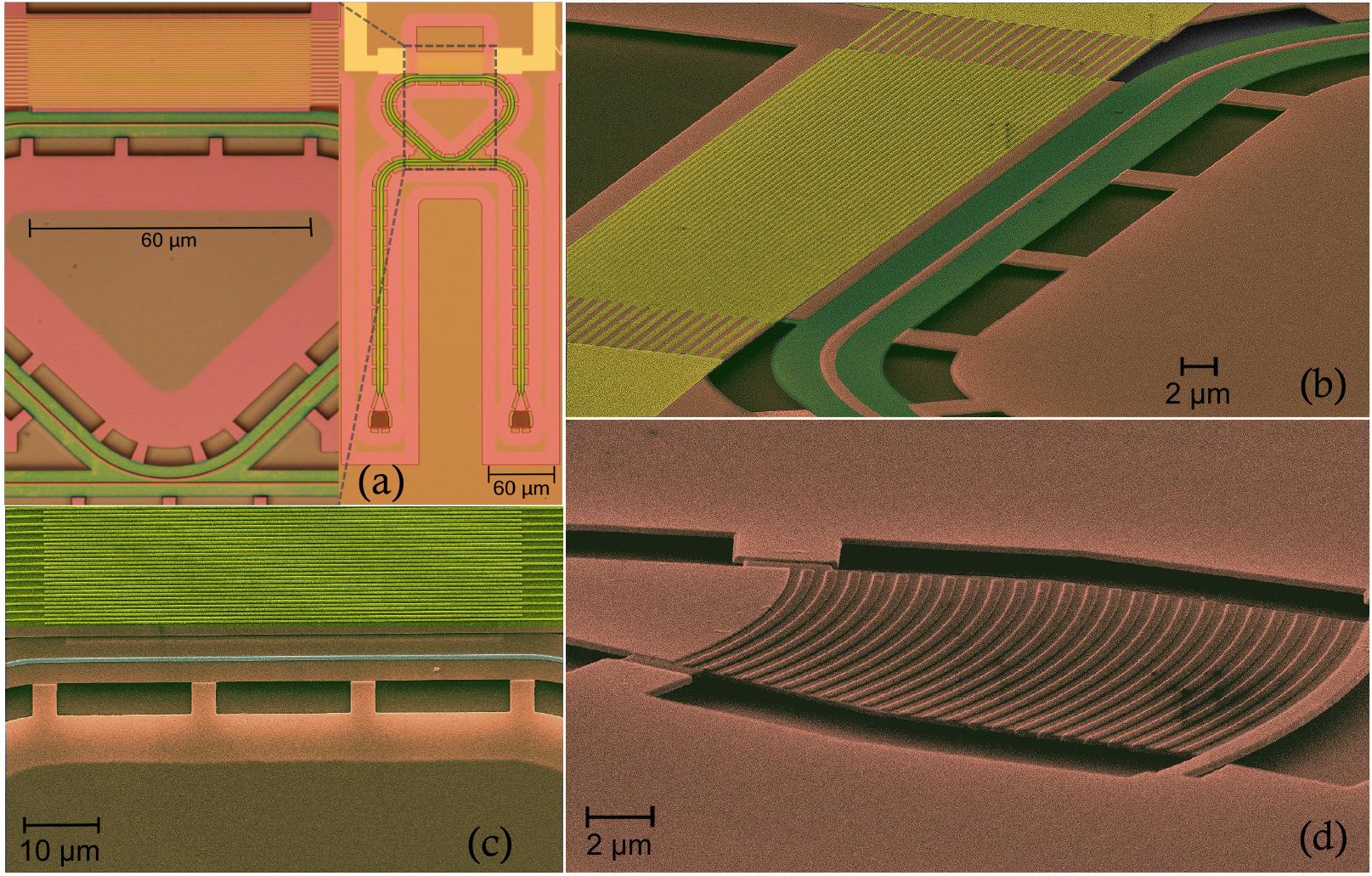}
\caption{Suspended 220 nm GaAs PIC platform with integrated Lamb wave resonators for piezo-optomechanical signal transduction (a) Optical microscope image of a representative device with zoomed in section of the microring resonator in conjunction with the IDT (Cr/Au). Suspended grating couplers, seen at the bottom-right are used to couple light into a rib waveguide. The rib waveguide is side-coupled to a triangular micro-ring resonator (coupling region shown in the zoomed-inset), which is also designed around a rib waveguide geometry. The Lamb wave resonator is defined by the suspended GaAs membrane under the IDT (gold fingers). (b),(c) False-colour SEM images showing the interconnection between the suspended ring resonator and the suspended IDT. The partially etched rib region (green) connects the Lamb wave resonator with the base of the triangular micro-ring resonator, allowing mode hybridization between the Lamb wave resonator mode and the breathing mode of a rib waveguide. The microring geometry enables resonant enhancement of the acousto-optic interaction, with the primary source of mechanical loss being the tethers (in (b) and (c)) used to hold the suspended rib waveguide/microring from the side. (d) False-colour SEM image of the partially etched suspended grating coupler.}
\label{fig:SEM}
\end{figure*}

We would like to note here that while Lamb wave resonators have been demonstrated around integrated photonics platforms before \cite{ghosh2016laterally, shao2019microwave}, to the best of our knowledge, a general design principle has not been outlined, and experimentally demonstrated, on how to engineer these structures to achieve the high $\eta_{peak}$ needed for quantum transduction. The key insight behind the mechanical supermode approach is to start from a mode that has intrinsically high $g_{0}$ and hybridize it with the Lamb wave resonance. In contrast to 1D nanobeam optomechanical crystals which have a higher $g_0$, the rib waveguide geometry studied in this work provides two key advantages: straightforward integration into an integrated photonics platform \cite{jiang2020suspended} because of relaxed dimensional tolerances and side-on coupling; and access to higher ($>$ 5 GHz) frequency mechanical modes with moderate $g_0$. This becomes critical for engineering quantum transduction \cite{balram2022piezoelectric} in a material with strong acousto-optic interactions, but relatively low speed of sound like GaAs ($v_{\scriptscriptstyle\text{SAW}} <$ \SI{3000}{\metre/\second}). Traditional 1D nanobeam optomechanical crystals, with $g_0\approx$ \SI{1}{\mega\hertz} have breathing mode frequencies $<$ 3 GHz, with minimum feature sizes $<$ \SI{100}{\nano\meter} \cite{balram2014moving}. Pushing these designs to mechanical frequencies $>$ 5 GHz, in order to achieve the requisite spectral separation from the strong optical pump \cite{balram2022piezoelectric}, places prohibitive constraints on nanofabrication.   

\vspace{-0.8em}
\section{\label{sec:level1}Device Architecture and characterization}
\vspace{-0.4em}
Fig.\ref{fig:SEM}(a) shows an optical microscope image of our suspended GaAs PIC platform with an integrated Lamb wave resonator for piezo-optomechanical signal transduction. Figs.\ref{fig:SEM}(b-d) show zoomed-in SEM images of the different sections of the device, including the grating coupler, rib waveguide, microring resonator and the Lamb wave resonator interfaced with the ring. The devices are fabricated on a \SI{220}{\nano\meter} thick GaAs device layer in a process flow derived from our previous work \cite{jiang2020suspended} on building suspended GaAs PIC platforms derived from a silicon photonics foundry process. The main changes to the process flow are the addition of the Cr/Au interdigitated transducers (IDT) (Figs.\ref{fig:SEM}(b,c)) to excite the acoustic waves. The presence of the metal layers modifies our cleaning procedure after the GaAs layer is suspended by undercutting the underlying AlGaAs layer. The complete fabrication process is detailed in Appendix \ref{sec:fab_process}.

\begin{figure}[t]
\includegraphics[scale=0.5]{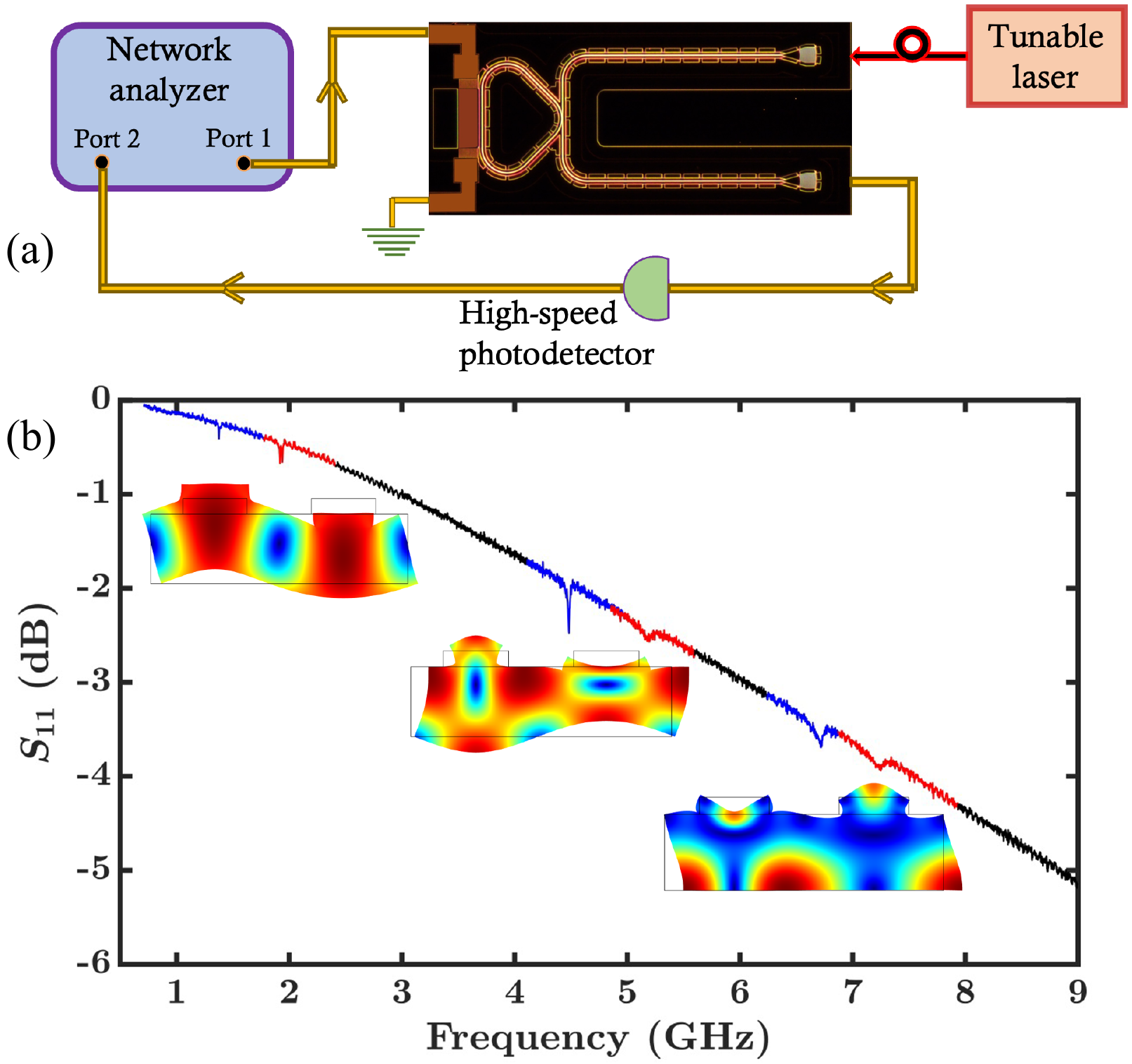}
\caption{(a) Schematic of the acousto-optic (AO) characterization setup used in the experiments. A vector network analyzer is used to drive the Lamb wave resonator (Port 1) and the AO modulation of the microring resonator is detected in transmission by a fast photodetector, whose output is connected to port 2 of the VNA (b) RF reflection ($S_{11}$ spectra for Lamb wave resonators with IDT period of \SI{0.8}{\micro\meter} and \SI{1}{\micro\meter}. The resonances corresponding to the $A_0$, $S_0$ and $A_1$ modes (see text) are highlighted by blue and red for the two IDTs respectively, and the background is shown in black. The mode shapes corresponding to the dips are shown in the inset.}
\label{fig:setupS11}
\end{figure}

The Lamb wave resonator is built to incorporate the rib waveguide. A schematic of the device cross-section, showing the interface between the Lamb wave resonator and the rib waveguide, and the resulting mode hybridization, is shown in Fig.\ref{fig:eigenmodes}(g) for reference. The rib waveguide is wrapped into a microring resonator as shown in Fig.\ref{fig:SEM}(a) to enhance the opto-mechanical interaction. The triangular shape of the microring is chosen in order to maximize the fraction of the ring that is modulated by the Lamb wave resonator. A schematic of the experimental setup used for coherent acousto-optic characterization is shown in Fig.\ref{fig:setupS11}(a). Light from a tunable telecom wavelength laser, Santec TSL-550, is coupled in through the suspended grating couplers (shown in Fig.\ref{fig:SEM}(d)) and routed through the on-chip suspended rib waveguides. The microring resonator is side-coupled to the rib waveguide with a coupling gap of \SI{350}{\nano\meter} and has an effective coupling length of \SI{60}{\micro\meter} with the lamb wave resonator. The microring has a total length of \SI{300}{\micro\meter}, which gives us an acousto-optic participation ratio ($\eta_\text{cav}$) of $20\%$. An optical transmission spectrum through a representative ring resonator device is shown in Fig.\ref{fig:AOmod} (blue curve). The nominal insertion loss in our grating couplers is $\approx$ 7 dBm per coupler and our microring resonators have nominal quality factors ($Q_o$) of $\approx$ 20,000. Our fiber array is kept some distance away from the grating couplers to avoid causing collapse of the suspended device. This results in an additional loss of $\approx$ 4 dBm between the fiber array and the air on each side, estimated by the difference in measured transmission with the fiber array hovering and close to landing on the coupler. The optical power drawn from the laser is 7 dBm (5 mW) in our experiments, which results in $\approx$ -4 dBm (0.398 mW) of power at the input of the microring cavity (used to estimate the intracavity circulating power in the experiments below) and $\approx$ -15 dBm (0.03 mW) of power received at the fast photodiode.

The microwave signal is coupled to the IDTs using standard RF probes (Picoprobe 40A GSG). The Lamb wave resonances of the membrane can be measured from the RF reflection ($S_{11}$) spectrum of the device. Representative $S_{11}$ spectra for devices with IDT periods \SI{1}{\micro\meter}  and \SI{0.8}{\micro\meter} are shown in Fig.\ref{fig:setupS11}(b) with their respective resonances highlighted in blue and red, respectively. The three dips correspond to the fundamental anti-symmetric (A$_{0}$), fundamental symmetric (S$_{0}$) and first-order anti-symmetric (A$_{1}$) Lamb wave modes, respectively \cite{zou2014high}. Finite element (FEM) simulation of the respective modal displacement profile is also depicted in the inset of Fig.\ref{fig:setupS11}(b). The nominal acoustic velocity of these three Lamb wave (plate) modes are \SI{1400}{\meter/\second}, \SI{4100}{\meter/\second} and \SI{5800}{\meter/\second}, compared to the standard SAW velocity of $\approx$ \SI{2800}{\meter/\second} in GaAs, and enable transduction of higher frequency optomechanical modes.

\section{Piezo-optomechanical transduction through Lamb wave resonators}

The acousto-optic (AO) interaction inside the waveguide can be accurately modelled using an FEM simulation (COMSOL Multiphysics) of the acoustic and optical fields and calculating the AO perturbation using an overlap integral between the two. In general, the AO interaction has two contributions of different physical origin. When the acoustic wave perturbs the rib waveguide, it changes the waveguide dimensions perturbing the overall optical field distribution. This is referred to as the moving boundary perturbation \cite{balram2014moving} and can be quantified using a surface integral over the waveguide boundaries of the optical and mechanical fields \cite{wiederhecker2019brillouin}, given in equation (\ref{eqn:mb}). In addition, the strain developed in the waveguide modifies the local refractive index due to the stress-optical (photoelastic) effect \cite{balram2014moving}. The volume integral between the optical and mechanical modes to estimate the photoelastic perturbation to the refractive index is quantified as equation (\ref{eqn:pe}).

\begin{eqnarray}
G_{\mathrm{MB}} \:&&= -\frac{\omega_0}{2} \frac{\oint \left(\text{Q}.\boldsymbol{\hat{\mathrm{n}}}\right) \left(|\text{E}_{\|}|^2\Delta \epsilon - \Delta \epsilon^{-1} |\text{D}_{\bot}|^2 \right) dS}{\int \epsilon |\text{E}|^2 dV} \label{eqn:mb} \\
G_{\mathrm{PE}} \:&&= \frac{\epsilon_{0} \omega_{0} n^{4}}{2} \frac{\int \! dV \! \begin{pmatrix} \text{E}_{x}^{\ast} \!\!\!& \text{E}_{y}^{\ast} \!\!\!& \text{E}_{z}^{\ast}\end{pmatrix}\!\!\begin{pmatrix} pS_1 & pS_6 & pS_5 \\ pS_6 & pS_2 & pS_4 \\ pS_5 & pS_4 & pS_3\end{pmatrix}\!\!\! \begin{pmatrix} \text{E}_x \\ \text{E}_y \\ \text{E}_z \end{pmatrix}}{\int \epsilon |\text{E}|^2 dV} \label{eqn:pe} \nonumber \\
\end{eqnarray} 

where $G_{\mathrm{MB}}$ and $G_{\mathrm{PE}}$ represent the gradient of the optical frequency with displacement ($d\omega/dx$) for moving-boundary and photoelastic contributions, respectively, and $\omega_0$ is the resonance frequency of the unperturbed optical mode. In equation (\ref{eqn:mb}), \text{Q} is the normalised displacement induced at the waveguide boundary, $\boldsymbol{\hat{\mathrm{n}}}$ is the normal vector to the surface boundary, $\text{E}_{\parallel}$ and $\text{D}_{\bot}$ are the tangential optical electric field and normal electric displacement field components respectively, that are continuous across the boundary, and $\epsilon$ denotes the optical permittivity of GaAs, where $\Delta \epsilon = \epsilon_{\mathrm{GaAs}} - \epsilon_\mathrm{{air}}$ and $\Delta \epsilon^{-1} = \epsilon^{-1}_{\mathrm{GaAs}} - \epsilon^{-1}_\mathrm{{air}}$. In equation (\ref{eqn:pe}), $\epsilon_0$ is the vacuum permittivity, $n$ is the refractive index of the medium (GaAs), $\text{E}$ represents the electric field components of the optical mode and $pS_m$ (m = 1-6) are the changes in the optical indicatrix coefficients due to strain $s_m$ (m = 1-6), defined as:
\begin{equation}
    \begin{pmatrix} pS_1\\pS_2\\pS_3\\pS_4\\pS_5\\pS_6 \end{pmatrix} = \begin{pmatrix}
    p_{11} & p_{12} & p_{12} & 0 & 0 & 0 \\
    p_{12} & p_{11} & p_{12} & 0 & 0 & 0 \\
    p_{12} & p_{12} & p_{11} & 0 & 0 & 0 \\
    0 & 0 & 0 & p_{44} & 0 & 0 \\
    0 & 0 & 0 & 0 & p_{44} & 0 \\
    0 & 0 & 0 & 0 & 0 & p_{44} 
    \end{pmatrix}
    \begin{pmatrix}
    s_1\\s_2\\s_3\\s_4\\s_5\\s_6
    \end{pmatrix}
    \label{eqn:indicatrix}
\end{equation}

where $p_{11}=-0.165$, $p_{12}=-0.14$ and $p_{44}=-0.072$ are the photoelastic coefficients of GaAs. The photoelastic matrix in equation (\ref{eqn:indicatrix}) is rotated to match the crystal orientation in our experiments, with the acoustic waves travelling along the $\left<110\right>$ direction.

For the experiments, the more relevant quantity to define is the vacuum optomechanical coupling rate $g_0$, which quantifies the interaction strength between a single photon and a single phonon in the mechanical mode (due to vacuum fluctuations). This is defined as $g_0 = Gx_{\scriptscriptstyle{\text{ZPF}}}$, where $x_{\scriptscriptstyle{\text{ZPF}}}$ is the zero-point mechanical fluctuation amplitude:
\begin{eqnarray}
    x_{\scriptscriptstyle{\text{ZPF}}} \;&&= \sqrt{\frac{\hbar}{2 m_{\text{eff}} \Omega_m}} \label{eqn:xzpf}\\ \nonumber \\
    \text{where,} \quad m_{\text{eff}}\;&&= \int_R \!\rho \: Q(r)^2 dV / \max_{R} \left(Q(r)^2\right) \label{eqn:meff}
\end{eqnarray}

is the effective mass of the phonon mode. The mode displacement is integrated over a simulation domain region  $R$, defined by a coordinate variable $r \in R$. $\rho$ is the density of the material (GaAs, isotropic) and Q is the mechanical displacement of the mode as define above in equation (\ref{eqn:mb}). 

In this work, we perform FEM simulations and overlap integrals of propagating modes across the device cross-section in 2D. Therefore, surface integral ($dS$) and volume integral ($dV$) in equations (\ref{eqn:mb}), (\ref{eqn:pe}) and (\ref{eqn:meff}) are replaced by $L_c \times dr$ (boundary integral) and $L_c \times dS$ (surface integral) respectively, where, $L_c$ is the length perpendicular to the simulation cross-section, also known as coupling length between the acoustic and the optical resonator (in this work, $L_c =$ \SI{60}{\micro\meter}). This also necessitates approximating the optical modes of a rib waveguide microring resonator with that of a straight rib waveguide, but given the relatively large bend radii in our devices, the mode perturbation due to curvature is small. In addition, the presence of the tethers will perturb the mechanical modeshape in 3D  \kcb{(see Appendix F, Fig.11(b))} but to make the computations tractable we rely primarily on the approximate 2D calculation of the acousto-optic overlap in this work. \kcb{The presence of the supermode can be inferred in full 3D simulations of a scaled version of the device, shown in Appendix F. A scaled version of the device is chosen mainly to fit the simulations in memory, and shows good correspondence between the 3D modeshapes and the 2D simulation results shown in Fig.1(g).}

The total optomechanical interaction strength ($g_{\text{0,tot}}$) is calculated as $(g_{\text{0,MB}} + g_{\text{0,PE}})* \eta_{\text{cav}}$, where $\eta_{\text{cav}}$ is the ratio of the rib waveguide section coupled to acoustic resonator ($L_c$ ), to that of the overall length of the microring. This acousto-optic participation ratio ($\eta_{\mathrm{cav}}$) in our current designs is $20\%$. The numerically evaluated $g_{0}$ from equations (\ref{eqn:mb}) and (\ref{eqn:pe}) has units of angular frequency (rad.Hz) but it is divided by $2\pi$ and expressed in Hz in this work. 

The acoustic and optical modes are simulated independently, as shown in Fig.\ref{fig:eigenmodes}, and the interaction strength is calculated by solving the overlap integrals described above. In order to find the peak interaction strength in this architecture, we first simulate the optical and the acoustic modes for the strip and the rib waveguides without the IDT section. This way we can also identify the shear horizontal (SH) breathing mode frequencies of the respective waveguides. For a strip waveguide (rib height = 0), we obtain two breathing modes at \SI{4.2}{\giga\hertz} (SH) and \SI{10.7}{\giga\hertz}, both of which result in $g_0$ higher than \SI{200}{\kilo\hertz}. The X and Y displacements of \SI{4.2}{\giga\hertz} mode are shown in Fig.\ref{fig:eigenmodes}(c,d), respectively. Although the strip waveguide geometry is potentially a good candidate for travelling wave microwave-to-optical transduction, it runs into the same problem as 1D photonic crystals, where the breathing mode cannot be efficiently excited by acoustic waves and $\eta_{\scriptscriptstyle\text{PIE}}\ll$ 1 \cite{balram2022piezoelectric}. Hence, the Lamb wave resonator is designed to work with a rib waveguide to create a mechanical supermode. We would like to note here that to achieve a strong acousto-optic interaction using this approach, it is critical that the rib displacement be of the right mode shape, with the right strain profile in the waveguide to generate a moderately high $g_0$ with the fundamental transverse electric (TE) mode of the rib waveguide. 

\begin{figure}[htbp!]
\includegraphics[scale=0.5]{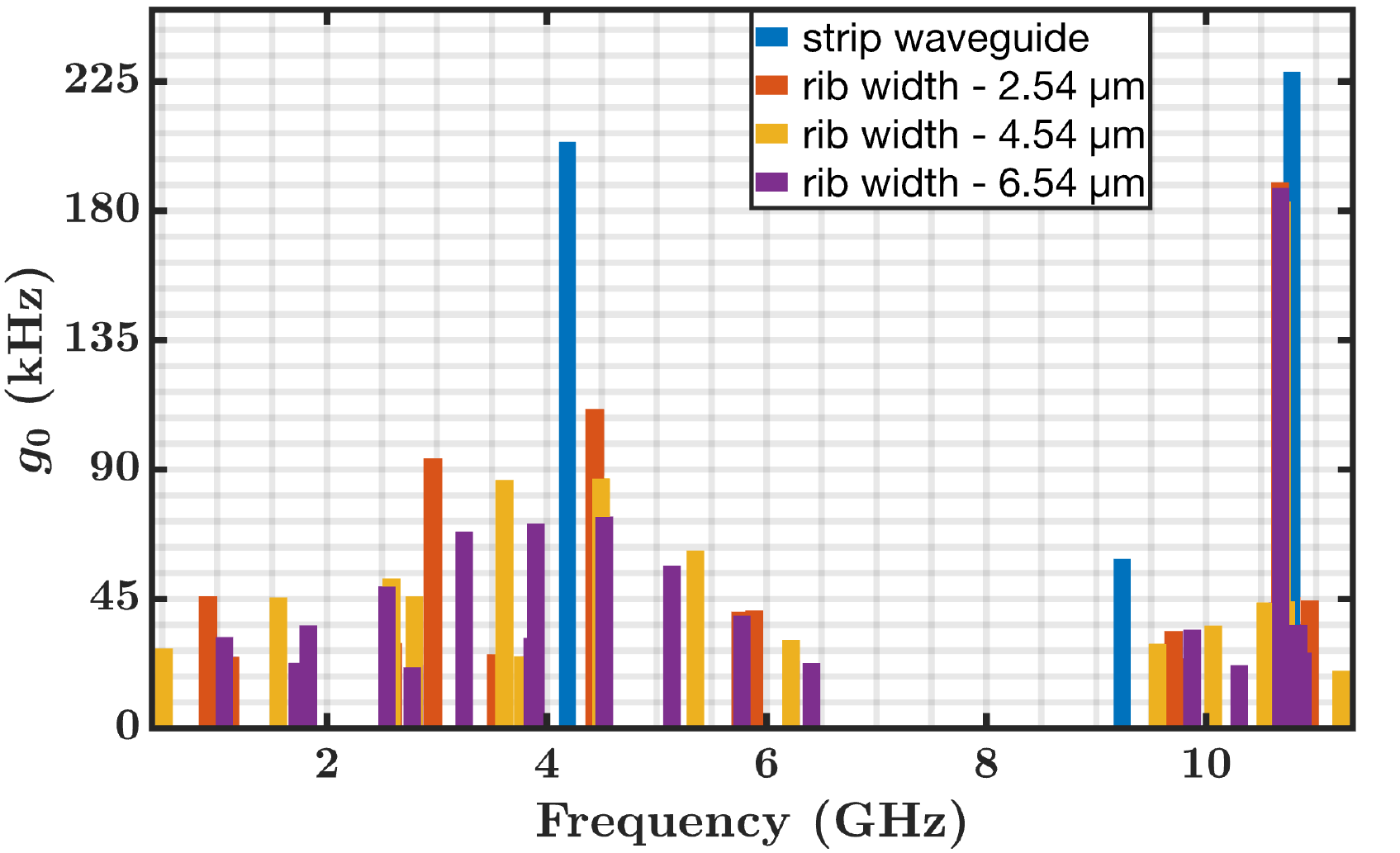}
\caption{Numerically estimated $g_0$ of the shear horizontal breathing modes  ($\approx$ 4.2 and 10.7 GHz) of a strip waveguide ($w$ = 540 nm, blue bars) and rib waveguides with rib widths of \SI{2.54}{\micro\meter}, \SI{4.54}{\micro\meter} and \SI{6.54}{\micro\meter}. Adding the rib sections split the original SH mode into a multiplet of modes with an accompanying reduction in $g_0$ due to mechanical mode delocalization in the ribs. In turn, the mode delocalization enables efficient coupling from a Lamb wave resonator through mode hybridization and a mechanical supermode formation.}
\label{fig:g0eigen}
\end{figure}

Fig.\ref{fig:g0eigen} depicts the numerically calculated $g_0$ values (in Hz) obtained for strip and rib waveguides of width \SI{540}{\nano\meter} and rib widths of \SI{2.54}{\micro\meter}, \SI{4.54}{\micro\meter} and \SI{6.54}{\micro\meter}. As we move from the strip to the rib waveguide geometry and the rib width increases, the shear horizontal mode splits into other modes and the $g_0$ value decreases due to mechanical mode delocalization, as can be seen in Fig.\ref{fig:g0eigen}. This is a fundamental tradeoff in the supermode approach in that the increase in $\eta_{\scriptscriptstyle\text{PIE}}$ comes at the cost of reduced $g_0$. The $g_0$ for the breathing mode reduces from $\approx$ 200 kHz for the strip waveguide to $\approx$ 35 kHz ($\approx$ \SI{1.9}{\giga\hertz}) for the rib waveguide with rib width \SI{6.54}{\micro\meter}, and to $\approx$ 25 kHz for the hybridized mode including the IDT. The X and Y displacements in the rib geometry for the shear horizontal mode ($\sim$ \SI{4.5}{\giga\hertz}) are shown in Fig.\ref{fig:eigenmodes}(e,f). The shear horizontal mode (\SI{4.5}{\giga\hertz}) is of prime interest in this work, despite \SI{10.7}{\giga\hertz} mode having a higher $g_0$, because most microwave qubits lie in the frequency range 3 - 7 \SI{}{\giga\hertz} \cite{bardin2021microwaves} and the \SI{10.7}{\giga\hertz} mode places prohibitive constraints on the IDT fabrication process. The breathing mode at \SI{10.7}{\giga\hertz} can be moved to $\approx$ \SI{7}{\giga\hertz} if the GaAs thickness is increased to $\approx$ \SI{340}{\nano\meter}.
 
In GaAs, the AO interaction is primarily dominated by the photoelastic effect. For reference, in the case of a rib waveguide with width 540 nm, thickness 220 nm, rib thickness 100 nm and rib width of \SI{3.54}{\micro\meter}, we calculate $g_{\mathrm{0,MB}}$ = \SI{3.12}{\kilo\hertz} and $g_{\mathrm{0,PE}}$ = \SI{-100.06}{\kilo\hertz}. As noted above, the shear horizontal breathing mode of a rib waveguide has a large overlap with the fundamental TE optical mode, leading to $g_{0}\approx$ 100 kHz. We would like to note again that all $g_0$ values are calculated for an overlap length of \SI{60}{\micro\meter}, which corresponds to the width of the fabricated Lamb wave resonator devices.

\begin{figure}[b]
\includegraphics[scale=0.46]{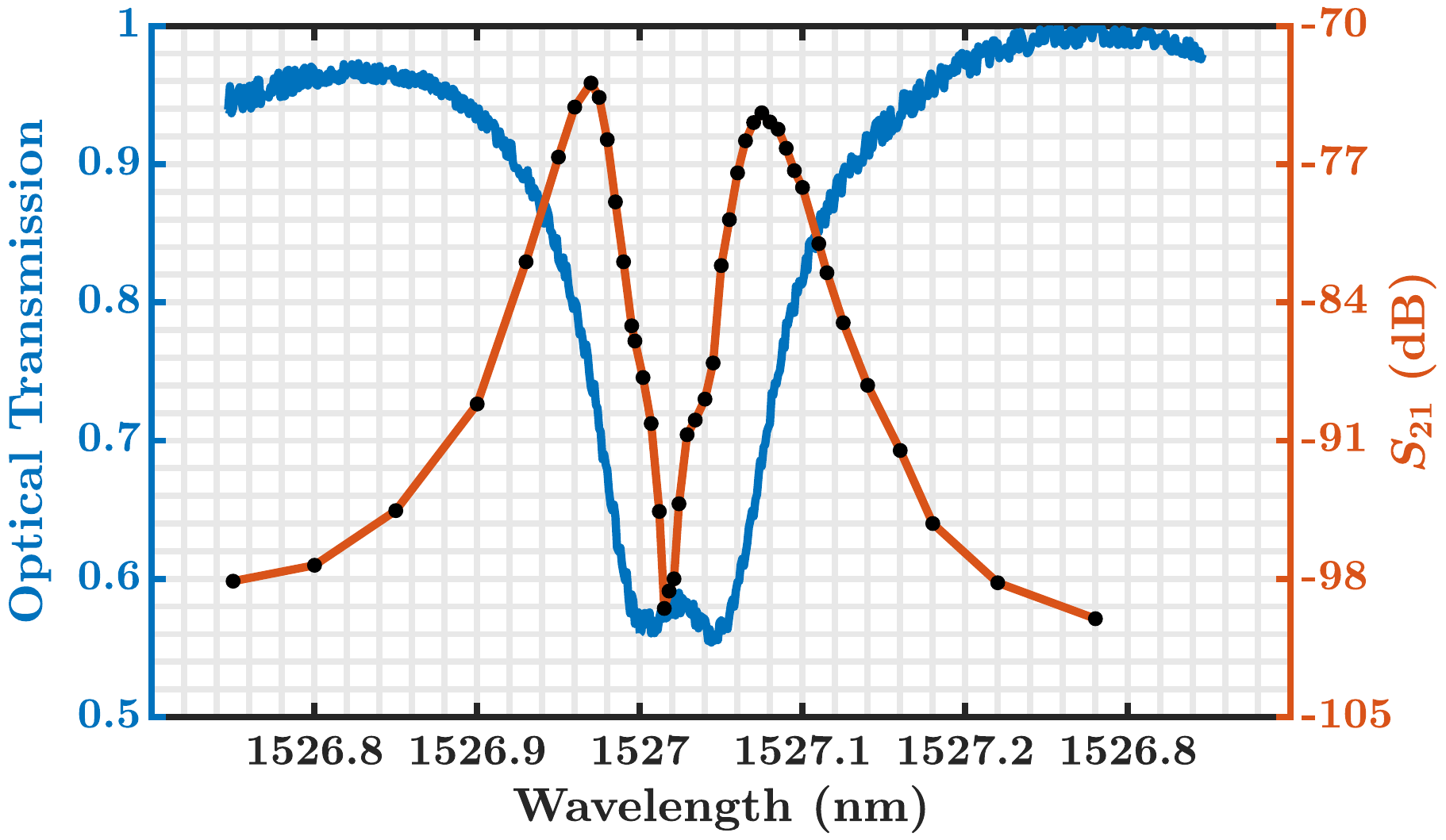}
\caption{Acousto-optic (AO) modulation for one of the mechanical modes ($f_m\approx$ \SI{2}{\giga\hertz} in Fig.\ref{fig:g0_S21}) as the telecom laser is tuned across one of the microring cavity resonances (blue, $Q_o\approx$ 20,000). The optical cavity lineshape converts the AO phase modulation induced by the mechanics into amplitude modulation (AM, black dots correspond to the experimental measurements), and it can be seen that the AM is peaked when the laser ($\omega_L$) detuning ($\Delta = \omega_L-\omega_0$) from the cavity centre ($\omega_0$) is $\Delta\approx\pm\kappa$/2, and is minimized when $\Delta\approx$ 0.}
\label{fig:AOmod}
\end{figure}

\begin{figure*}[htbp]
\includegraphics[scale=0.98]{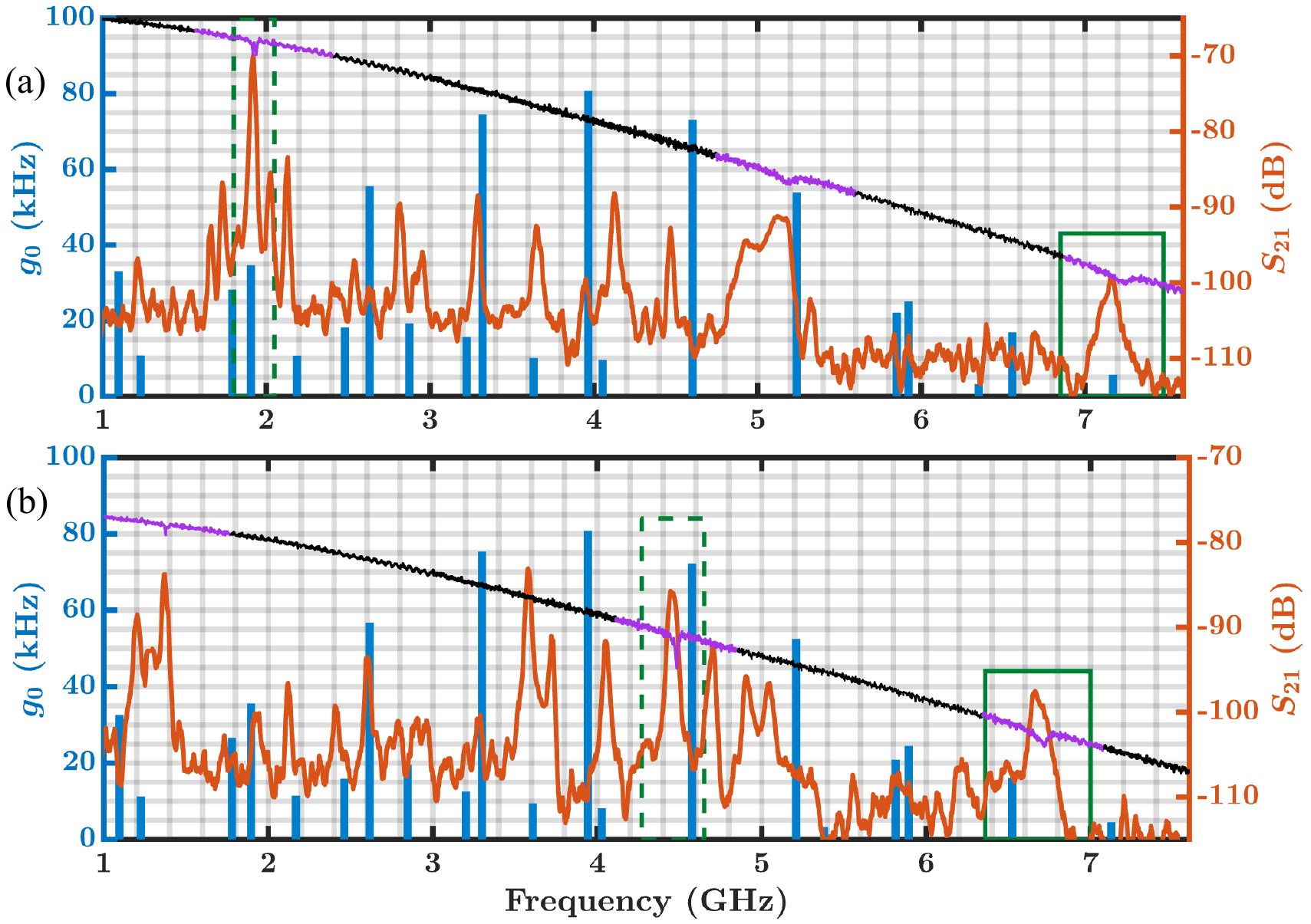}
\caption{Coherent AO modulation spectrum ($S_{21}$, dark orange) of two microrings driven by integrated Lamb wave resonators of period \SI{0.8}{\micro\meter} (a) and \SI{1}{\micro\meter} (b) respectively. The $S_{11}$ response of the transducers is shown in black and the resonant modes are highlighted in purple. The numerically estimated $g_0$ of various mechanical modes for the rib waveguide are shown by the blue bars, where the width of the bars is \SI{50}{\mega\hertz}, representing the deviation of $\pm$\SI{25}{\mega\hertz} with an error of $\pm$\SI{25}{\nano\meter} in the estimation of the rib widths from the SEM imaging. The measurements and numerical calculations have been made for fabricated devices with (a) waveguide width of \SI{0.564}{\micro\meter} and rib width of \SI{6.378}{\micro\meter}, and (b) waveguide width of \SI{0.574}{\micro\meter} and rib width of \SI{6.408}{\micro\meter}, as shown by the SEM images in Appendix \ref{sec:width}. When the $S_{11}$ response of the device lines up with one of the high $g_0$ modes (shown by the green dashed box in (a)), there is a significant increase in transduction efficiency due to supermode formation. On the other hand, when the modes are mis-aligned (indicated by the dashed box in (b)), the transduction is inefficient, even if more energy is coupled into the Lamb wave resonator (note the size of the $S_{11}$ dip in the two shaded green boxes). Lamb wave resonators also enable higher frequency ($\approx$ 7 GHz) piezo-optomechanical transduction, compared to standard surface acoustic wave devices, as indicated by solid green boxes in (a) and (b).}
\label{fig:g0_S21}
\protect\tikz[remember picture] \draw[overlay, red] ([yshift=0.25em]pic cs:cap1s) -- ([yshift=0.25em]pic cs:cap1e);
\tikz[remember picture] \draw[overlay, red] ([yshift=0.25em]pic cs:cap2s) to ([yshift=0.25em]pic cs:cap2e);
\end{figure*}

To observe the coherent AO modulation, we use the experimental setup shown in Fig.\ref{fig:setupS11}. The laser is parked on the shoulder (the detuning $\Delta\approx{\kappa}/2$, where $\kappa$ is the optical cavity linewidth) of the optical cavity resonance and the Lamb wave resonator is driven from port 1 of the vector network analyzer (R\&S-ZVL 13). The transmitted optical signal is detected using a high speed photodetector (Thorlabs DXM30BF), whose output is connected to port 2 of the VNA. This performs a coherent acousto-optic $S_{21}$ measurement \cite{balram2016coherent}. The optical cavity converts the phase modulation induced by the AO interaction into an amplitude modulation that can be detected using the high-speed photodetector. This can be seen experimentally in Fig.\ref{fig:AOmod}, where the AO modulation ($S_{21}$) for a representative mechanical mode ($f_m\approx$ \SI{2}{\giga\hertz}, in Fig.\ref{fig:g0_S21}) is shown as the laser is tuned across the optical cavity resonance. It can be clearly seen that the modulation is peaked, when the laser is tuned to the point of maximum slope of the cavity lineshape ($\Delta\approx\kappa/2$) and is minimized at the cavity centre ($\Delta\approx0$).

By parking the laser at the optimal detuning ($\Delta\approx\kappa/2$), and sweeping the VNA frequency, we can measure the coherent AO modulation induced by the Lamb wave resonator. Fig.\ref{fig:g0_S21}(a,b) shows the experimentally measured AO-$S_{21}$  spectra (in dark orange) for representative Lamb wave devices with IDT periods of \SI{0.8}{\micro\meter} (a) and \SI{1}{\micro\meter} (b) respectively, recorded with an output optical power of \SI{30}{\micro\watt} measured at the photodetector. The $S_{11}$ spectra of the devices, shown in Fig.\ref{fig:setupS11}(b), are also overlaid for reference in black, with the resonant modes highlighted in purple. There is a clear correspondence between the measured $S_{11}$ dips (purple) and the $S_{21}$ peaks (dark orange) at the same frequencies in both cases showing that Lamb wave resonators can effectively transduce mechanical motion in the microring. In addition, there are several other peaks that are also transduced, even though there is negligible RF energy converted into mechanical motion by the Lamb wave transducer. 

We can understand the observed behaviour in detail through FEM simulations. Fig.\ref{fig:g0_S21} also shows the simulated $g_0$ for a variety of mechanical modes of the fabricated rib waveguide, with rib width of $\approx$ \SI{6.378}{\micro\meter} for the device in Fig.\ref{fig:g0_S21}(a) and $\approx$ \SI{6.408}{\micro\meter} in  Fig.\ref{fig:g0_S21}(b). The waveguide widths in the two cases are \SI{0.564}{\micro\meter} and \SI{0.574}{\micro\meter} respectively. The device dimensions are estimated from SEM images of the device post fabrication, the images are shown in Appendix \ref{sec:width}. The overlap integrals are calculated for the mechanical mode with the fundamental TE optical mode. We can see that the additional peaks in the $S_{21}$ spectrum are approximately aligned (subject to fabrication tolerance and uncertainty in determining device dimensions) with high $g_0$ mechanical modes, which provide moderate transduction efficiencies, even though they are mis-aligned with the Lamb wave resonator frequency (cf. modes at $\approx$ 3.5 GHz in Fig.\ref{fig:g0_S21}(b)). To account for dimensional uncertainty, the widths of the blue bars in Fig.\ref{fig:g0_S21} is 50 MHz, corresponding to a dimensional uncertainty of $\pm$ 25 nm. From Fig.\ref{fig:g0_S21}(a), when the $S_{11}$ dips (purple) are exactly lined up with one of the high-$g_0$ modes of the rib waveguide ($f_m\approx$ 1.9 GHz, indicated by the green dashed box), there is a significant increase in the AO modulation ($S_{21}$) amplitude. This increase in overall transduction efficiency is a clear signature of the formation of a mechanical supermode between the Lamb wave resonator and the rib waveguide breathing mode at $\approx$ 1.9 GHz. 
\tikz[remember picture] \draw[overlay, red] ([yshift=.25em]pic cs:ribwstart) -- ([yshift=.25em]pic cs:ribwend);

\begin{figure}[t]
\includegraphics[scale=0.49]{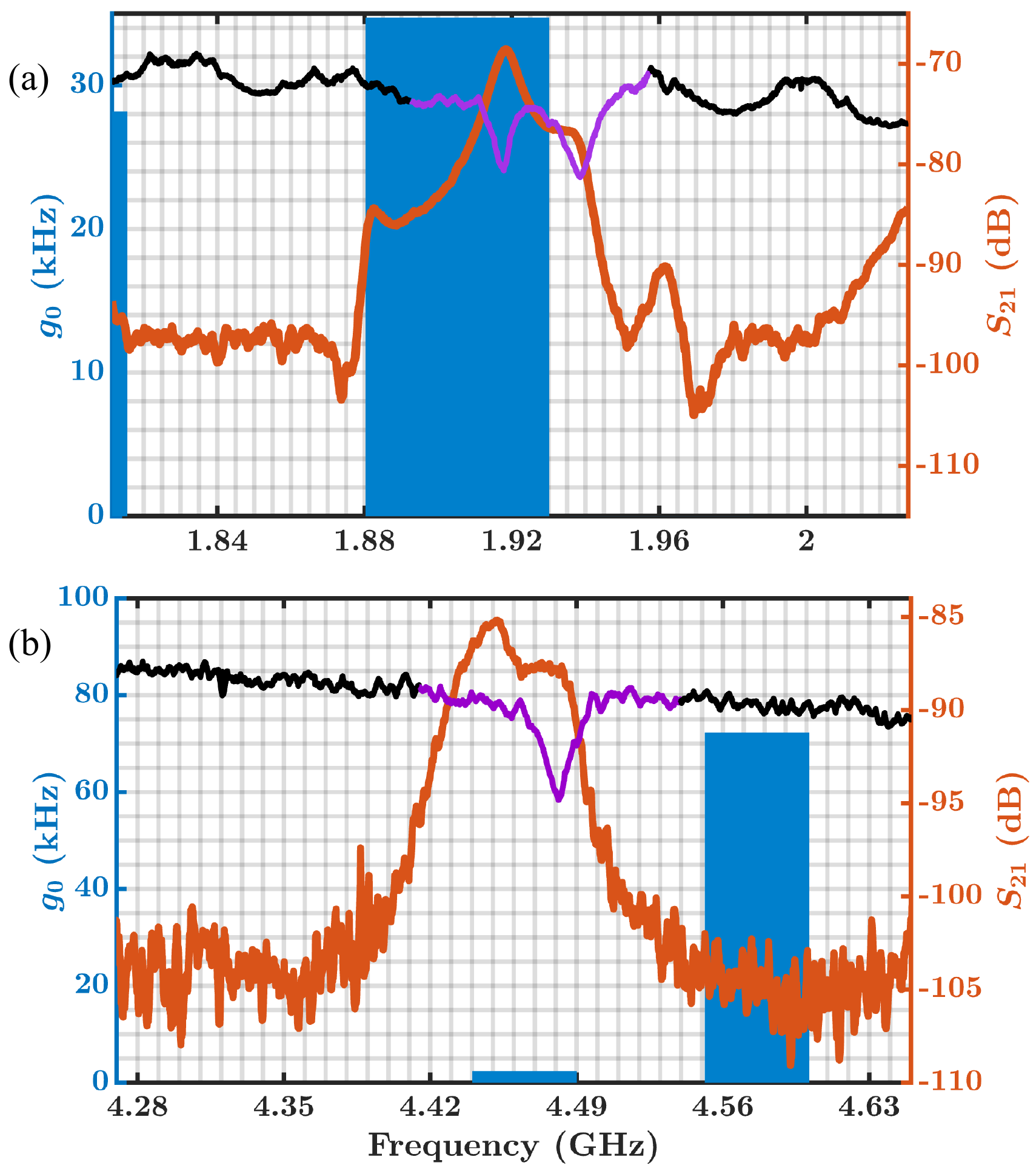}
\caption{Zoomed-in AO-$S_{21}$ spectra (dark orange) from the dashed green boxes in Fig.\ref{fig:g0_S21} for the Lamb wave resonators with \SI{0.8}{\micro\meter} (a) and \SI{1}{\micro\meter} (b) periods respectively. In (a), when the $S_{11}$ response of the Lamb wave resonator exactly lines up with a high-$g_0$ mode of the rib waveguide, the transduction efficiency ($S_{21}$) is significantly enhanced. This enhancement has been observed for multiple devices with same design parameters, as shown by the plot in Appendix \ref{sec:3devices}. On the other hand, when the two are offset, $\eta_{peak}$ is reduced. Note in (b) that the $S_{11}$ of the second IDT is significantly higher (refer to Table \ref{tab1} for $S_{11}$ magnitudes), but the $S_{21}$ is $\approx$ \SI{14}{\decibel} lower. In (b), one can clearly see that the $S_{21}$ is a doublet, where the first peak is due to a mechanical mode of the waveguide at $\approx$ \SI{4.45}{\giga\hertz}, see text for details. Note that the bar width (\SI{50}{\mega\hertz}) is the deviation in frequency of the mechanical mode (numerically calculated) for $\pm$\SI{25}{\nano\meter} variation in waveguide rib width.}
\label{fig:g0_S21_zoom}
\end{figure}

Fig.\ref{fig:g0_S21_zoom} (a) and (b) provide a zoom-in spectrum of the two dashed green boxes in Fig.\ref{fig:g0_S21}. As can be seen from Fig.\ref{fig:g0_S21_zoom}(a), when the $S_{11}$ dip lines up with a moderately high $g_0$ mode of the rib waveguide (blue bar), the transduction efficiency ($S_{21}$ magnitude) is significantly enhanced. For reference, in Fig.\ref{fig:g0_S21_zoom}(b), the $S_{11}$ response is larger, showing that more of the input microwave energy is coupled into acoustic vibration in the resonator. On the other hand, the spectral misalignment between the $S_{11}$ resonance and a mechanical mode of the waveguide with moderate $g_0$ (blue bars) results in a significantly lower overall transduction efficiency. Fig.\ref{fig:g0_S21_zoom}(b) also provides additional evidence for the importance of lining up the spectral responses. The $S_{21}$ response is a doublet with the second peak exactly lined up with the IDT response (in purple). The first peak lies just off the shoulder of the IDT response and corresponds to a moderate $g_0$ mode ($\approx$ 5 kHz) of the waveguide. It is clear from the relative sizes of the two peaks that even relatively poor alignment with a mechanical mode of the waveguide results in a significantly higher ($\approx$ 4 dB) transduction efficiency, even though most of the microwave power is reflected off the Lamb wave resonator. We were unable to see the mechanical mode signatures in the thermal noise spectra of the microrings due to the sub-optimal $Q_{o,m}$ of the bare optomechanical device, as discussed in the following section. Observing the peaks (corresponding to the blue bars in Fig.\ref{fig:g0_S21}) in the thermal noise spectrum would have provided us independent experimental verification of the supermode formation, without the need to rely on numerical simulations as in this work. On the other hand, the evidence provided in Figs.\ref{fig:g0_S21} and \ref{fig:g0_S21_zoom}, and most importantly, the significant increase in transduction efficiency which is the key parameter of interest, all point strongly to supermode formation. We acknowledge that there is an intrinsic uncertainty with assigning modes to peaks in Fig.\ref{fig:g0_S21}, given the uncertainty in geometry due to nanofabrication errors. On the other hand, we find the supermode formation relatively robust, provided the mode alignment occurs within the bandwidth of the IDT. In Appendix D, we attach measurements from additional devices designed to be nominally identical to the device in Fig.\ref{fig:g0_S21}(a). Both the doublet splitting in the $S_{11}$ and a high $\eta_{peak}$ can be observed in all cases.    

The suitability of the Lamb wave resonator geometry for achieving high frequency transduction can also be seen from the solid green boxes in Fig.\ref{fig:g0_S21}(a,b) which show modes $>$ 7 GHz being transduced by the Lamb wave resonator's $A_1$ mode. Although the overall transduction efficiency in this case is limited due to spectral misalignment, these are already the highest frequency mechanical modes that have been piezoelectrically transduced in a slow acoustic velocity material like GaAs. As mentioned above, by working with a 340 nm thick GaAs layer, one can in principle line up the high $g_0$ mode at 10.7 GHz in Fig.\ref{fig:g0eigen} with this IDT resonance to create an effective supermode and increase the overall $\eta_{peak}$.

\section{Analysis: signal transduction efficiency enhancement}

To quantify the overall signal transduction efficiency and in particular, the enhancement due to the supermode, we extract the experimental acousto-optic coupling strength $g_0$ and the overall signal transduction efficiency for different mechanical modes using the magnitude of the measured AO $S_{21}$ signal. To estimate this, we assume we are operating with weak microwave input power and in the regime where the mechanical frequency ($\Omega_m$) is comparable to but does not exceed the optical cavity decay rate ($\kappa$) \cite{aspelmeyer2014cavity, kippenberg2007cavity}. As long as the device is operating at weak microwave inputs with small optical power in the cavity and $\Omega_m \gg \gamma$, the relation between $S_{21}$ and $g_0$ is given by: 
\begin{eqnarray}
    S_{21}=\frac{32g_{0}^2\gamma_e\kappa_e^2R_{PD}^2P_{rec}^2}{\hbar\Omega_m\gamma^2\kappa^2R_{load}} \left[\frac{\Omega_m^2 + \kappa^2}{4\Omega_m^4 + \kappa^4}\right]
    \label{eqn:S21g0}
\end{eqnarray}

valid in both the sideband resolved and unresolved regimes. Here $\kappa$ ($\kappa_e$) and $\gamma$ ($\gamma_e$) are the total decay rate (external coupling rate) of the optical and acoustic mode respectively, $R_{PD}$ is the conversion gain of the photodetector (16.5 V/W), $P_{rec}$ is the power received at the photodetector and $R_{load} = 50 \Omega$ is the impedance of the input microwave source. The detailed analysis and derivation are provided in the Appendix \ref{sec:derivation}.

\begin{table}[htbp]
\caption{Experimental parameters and extracted transduction efficiency from the AO-$S_{21}$ data in Fig.\ref{fig:g0_S21}.}
\begin{tabular}{|c|c|c|c|c|c|c|}
     \hline 
     Parameters&\multicolumn{3}{c|}{\thead{IDT period \\ \SI{0.8}{\micro\meter}}} &\multicolumn{3}{c|}{\thead{IDT period \\ \SI{1}{\micro\meter}}} \\
     \hline \hline
     $\omega_0/2\pi$ (THz)&\multicolumn{3}{c|}{\thead{196.33}} &\multicolumn{3}{c|}{\thead{196.6}} \\
     \hline
     $Q_o$&\multicolumn{3}{c|}{\thead{20,150}} &\multicolumn{3}{c|}{\thead{19,861}} \\
     \hline
     $\kappa/2\pi$ (GHz)&\multicolumn{3}{c|}{\thead{9.74}} &\multicolumn{3}{c|}{\thead{9.9}} \\
     \hline
     $\kappa_e/\kappa$&\multicolumn{3}{c|}{\thead{0.13}} &\multicolumn{3}{c|}{\thead{0.19}} \\
     \hline
     Acoustic Mode&\thead{A$_0$} &\thead{S$_0$} &A$_1$ &A$_0$ &S$_0$ &A$_1$ \\
     \hline 
     $\Omega_m/2\pi$ (GHz) &1.94&5.18&7.21&1.38&4.48&6.72 \\
     \hline
     $Q_m$ &319.5&323.6&360.8&230&293.2&134.3 \\
     \hline
     $\gamma/2\pi$ (MHz) &6.07&16&20&6&15.3&50 \\
     \hline
     $\gamma_e/\gamma$ &0.013&0.0045&0.0045&0.012&0.025&0.011 \\
     \hline
     $S_{11}$ (dB) &-0.23&-0.08&-0.08&-0.21&-0.43&-0.2 \\
     \hline
     $S_{21}$ (dB) &-68.5&-89.8&-95.96&-81.1&-81.6&-94.8 \\
     \hline
     $g_0/2\pi$ (kHz) &4.37&1.75&1.34&0.62&1.17&0.96 \\
     \hline
     $N_{\mathrm{cav}}\times (10^4)$ &1.278&1.278&1.278&1.858&1.858&1.858 \\
     \hline
     \rowcolor{rose}[\tabcolsep][4pt]
     $\eta_{peak}\times (10^{-7})$ &1.17&0.027&0.01&0.045&0.15&0.012 \\
     \hline 
\end{tabular}
\label{tab1}
\end{table}

Table \ref{tab1} lists the experimental transduction efficiency and associated opto-electro-mechanical system parameters for the devices shown in Fig.\ref{fig:g0_S21}. In our simulations, we calculate $g_0/2\pi \approx$ 25 kHz for 1.94 GHz mode with Lamb wave hybridisation (resonator length = \SI{300}{\micro\meter} and overlap length = \SI{60}{\micro\meter}). This gives us $g_{0, \text{tot}}$ of 5 kHz for $\eta_\text{cav} = 0.2$, which is in reasonable agreement with experimentally calculated value of $g_0/2\pi$ of 2.18 kHz. The highest $\eta_{peak}$ measured in our system is $\approx1.17*10^{-7}$ for the 1.92 GHz ($A_0$) mode of the \SI{0.8}{\micro\meter} Lamb wave resonator, which forms the supermode as discussed above in Fig.\ref{fig:g0_S21}(a). To quantify the supermode enhancement factor, we can compare the $\eta_{peak}$ for the $A_0$ modes of the two devices (with periods 0.8 and \SI{1}{\micro\meter}), and find an enhancement factor of $\approx$ 25$\times$. We also find an enhancement of $\approx$ 8$\times$ over the $S_0$ mode of the \SI{1}{\micro\meter} Lamb wave resonator, even though the $S_{0}$ mode is far more effectively transduced by the IDT ($\gamma_e/\gamma\approx2 \times$ higher) and the $S_0$ mode is $\approx$ 2.3$\times$ higher in frequency. Our $\eta_{peak}$ is the highest demonstrated for III-V platforms, which are usually characterized by high $g_{0}$ and low $k^2_\text{eff}$ (see \cite{balram2022piezoelectric} for a table listing the performance metrics of other piezoelectric optomechanical platforms). While our overall $\eta_{peak}$ is below state of the art results in stronger piezoelectric (LiNbO$_3$) and hybrid (AlN/Si) platforms, below we discuss how our devices can be improved to increase $\eta_{peak}$ significantly in the near-term.

We would like to note here that while we have used the enhancement in $\eta_{peak}$ and the spectral alignment of modes to infer the formation of the supermode, ideally, one would like to map the acoustic displacement field \cite{zheng2020imaging} or tune the resonance frequency of the Lamb wave resonator in-situ (by local heating for instance) to observe the anti-crossing of the Lamb wave mode with the waveguide breathing mode. These measurements, along with the thermal noise spectra, as discussed above, are key to providing direct experimental evidence for the supermodes, without relying on numerical simulations which are sensitive to fabrication geometry. On the other hand, it is the overall $\eta_{peak}$ that is the main quantity of interest for microwave to optical signal transduction, and our work already shows an $\approx10^4$ enhancement over current GaAs devices \cite{balram2022piezoelectric}, with significant prospects for improvement as discussed below. We would also like to note that in Fig.\ref{fig:g0_S21_zoom}(a), the IDT $S_{11}$ shows a pronounced doublet splitting (for reference, compare the $S_{11}$ in Fig.\ref{fig:g0_S21_zoom}(b)), which is another indirect signature for supermode formation, but without an in-situ tuning mechanism, it is hard to quantify the coupling strength between the rib waveguide SH mode and the Lamb wave resonator experimentally.

\section{Outlook: Prospects for quantum transduction}

To achieve quantum transduction in this platform, the microwave to optical photon transduction efficiency, $\eta_{peak}\approx1$, which requires a significant improvement in current device performance. Although the supermode approach helps with increasing the $\eta_{peak}$ considerably in our current devices, the overall device performance is still limited by the bare device optical ($Q_o$) and mechanical ($Q_m$) quality factors. In particular, the $Q_o$ is $\approx 2*10^4$, which is significantly lower than one would expect from a rib waveguide microring geometry. It is possible that our devices have some residual underside roughness after the AlGaAs undercut, due to the modified cleaning procedure. We believe the main factor for optical loss in our devices is the limited thickness (\SI{1}{\micro\meter}) of bottom cladding (air following the undercut of AlGaAs). This is explored in more detail in Appendix \ref{sec:loss}. Improvements in the fabrication (cleaning) process, along with incorporation of alumina surface passivation \cite{najer2019gated}, and moving to GaAs wafer with higher AlGaAs thickness ($\geq$ \SI{1.5}{\micro\meter}) should improve our $Q_o$ significantly. 

The $Q_m$ is similarly $<$ 400 in our devices, mainly due to the tethering geometry used. As can be seen in Fig.\ref{fig:SEM}(c), there are four tethers that are holding the waveguide in our first-generation AO devices in this suspended GaAs platform (mainly due to the lack of tensile stress in GaAs) and each of them serves as a significant source of mechanical loss in our devices \kcb{(see Appendix F, Fig.11(b) for a 3D simulation showing mode leakage through the tethers)} . Moving the tethers to field nodes or removing them entirely should result in devices with $Q_m$ approaching standard Lamb wave resonators \cite{piazza2006piezoelectric}, with further $Q_m$ enhancements possible by incorporating more sophisticated bi-convex resonator geometries \cite{tu2016vhf}. A near-term goal is to understand the limit of $Q_m$ in these Lamb wave resonators at cryogenic temperatures. 

The biggest improvement in transduction efficiency comes from designing the supermode around the right optomechanical mode of the cavity, since $S_{21}{\propto}g_0^2$. This was the main motivation behind designing the supermodes around 1D nanobeam optomechanical crystals in the original proposal \cite{wu2020microwave} with $g_0>$ 1 MHz \cite{balram2014moving}, an improvement of 25$\times$ over the rib waveguide modes shown here. As discussed above, the key with ensuring high $\eta_{peak}$ using the supermode approach is to ensure that the $g_0$ doesn't drop too much due to mode hybridization. This is ultimately set by the size of the Lamb wave resonator needed for efficient RF-mechanical signal transduction and is one of the key challenges in working with a low piezoelectric coefficient platform like GaAs. Moving to a hybrid AlN-on-GaAs platform, and reducing the size of the Lamb wave resonator would allow us to maintain high $g_0$ and $\gamma_e$ simultaneously.

In the near-term, improving $Q_o$ by 10$\times$ using surface passivation and higher AlGaAs thickness, and achieving a 100$\times$ increase in $Q_m$ by better mechanical resonator design and cryogenic operation seem feasible. Incorporating an in-situ tuning mechanism to achieve a supermode between the SH breathing mode and the $S_0$ Lamb wave resonator mode, which results in higher piezoelectric transduction, would result in a further improvement. Assuming an $N_{cav}\approx10^4$ as in our current experiments, this gives us an $\eta_{peak}$ of $\approx10^{-5}-10^{-3}$, which we can realistically hope to achieve in the near future. The final step towards $\eta_{peak}\approx1$ will likely require a move towards a hybrid AlN-on-GaAs platform, mostly for shrinking the number of IDT finger pairs on the Lamb wave resonator front-end. We would like to reiterate that the supermode principle is independent of the underlying material platform used and can be used to enhance the transduction efficiency in any material platform. For instance, the ideas presented in this work can be applied to GaP, a material with similar piezoelectric and optomechanical device parameters as GaAs, but which provides better cryogenic (mK temperature range) performance due to reduced surface heating, especially in a 1D nanobeam optomechanical crystal geometry.

\section{Conclusions}

In summary, we have experimentally demonstrated that Lamb wave resonators, suitably hybridized with high $g_0$ optomechanical modes into mechanical supermodes, can result in a significantly enhanced microwave to optical transduction efficiency. While we have shown a 25$\times$ improvement in these proof-of-principle devices, our overall transduction efficiency is limited by the bare opto-mechanical device performance. Improving the bare device performance and engineering in-situ mode tuning to form supermodes around desired optomechanical modes sets the stage for achieving quantum transduction using this approach.

\section{Acknowledgements}

We would like to thank the European Research Council (ERC-StG, SBS3-5, 758843) and the Engineering and Physical Sciences Research Council (EP/V052179/1) for funding support. Nanofabrication was carried out using equipment funded by an EPSRC captial equipment grant (EP/N015126/1). The GaAs wafers were obtained from the EPSRC national epitaxy facility in Sheffield through a pump-prime award. We would like to thank Kartik Srinivasan, Marcelo Wu, Emil Zeuthen, Biswarup Guha, Seung-Bo Shim, Stefano Valle, Laurent Kling, Vinita Mittal, Robert Thomas, Rob Airey and Andy Murray for useful discussions and suggestions. AK would like to acknowledge funding from the Bristol Quantum Engineering Center for Doctoral Training, EPSRC-NPIF grant award EP/LO15730/1. 

\appendix \label{appendix}
\section{Extracting $g_0$ and transduction efficiency ($\eta$)} \label{sec:derivation}
\subsection{Input-output formalism}
The device can be modelled as a coupled optomechanical system with an acoustic resonator driven by a microwave signal and an optical resonator coupled evanescently to a waveguide. The coupling exists in the form of a feedback loop where the mechanical motion shifts the optical resonance frequency of the cavity which in turn exerts a radiation pressure force. The system can be analytically treated by using input-output formalism which provides us Heisenberg-Langevin equations of motion \cite{gardiner2004quantum,aspelmeyer2014cavity}, as given by,
\begin{eqnarray}
    \dot{a} =&&\; \left(i\Delta - \frac{\kappa}{2}\right)a - ig_{0}a\left(b + b^{\dagger}\right) + \sqrt{\kappa_e} a_{in} \label{eqn:HLa}\\
    \dot{b} =&&\; -\left(i\Omega_m + \frac{\gamma}{2}\right)b - ig_{0}a^{\dagger}a + \sqrt{\gamma_e} b_{in}\;, \label{eqn:HLb}
\end{eqnarray}
where $a$($a^{\dagger}$) and $b$($b^{\dagger}$) are the annihilation(creation) operators of the optical and acoustic modes, respectively, $\Omega_m$ is the resonance frequency of the acoustic resonator, $g_0$ is the single-photon optomechanical coupling strength, $\omega_L$ is the laser pump frequency, $\Delta = \omega_L - \omega_0$ is the laser detuning with respect to the optical resonance frequency $\omega_0$, $a_{in}$ and $b_{in}$ are the input optical and microwave fields, respectively, $\kappa = \kappa_e + \kappa_i$ and $\gamma = \gamma_e + \gamma_i$ are the total decay rates of the optical and acoustic fields, respectively, with subscript `$e$' denoting the external coupling rate and subscript `$i$' denotes the intrinsic loss. The decay rate $\kappa$ is related to the optical quality factor ($Q_{o}$) as $\omega_0/Q_{o}$ and similarly $\gamma = \Omega_m/Q_m$. The above equations are described with respect to a frame rotating with the laser frequency $\omega_L$, i.e., $a^{actual} = e^{-i\omega_L t}a^{above}$. The optical external coupling rate $\kappa_e$ can be calculated from the Lorentzian fitting of the optical transmission ($T$) \cite{xiang2020effects} as,
\begin{equation}
    \kappa_e = \frac{\kappa}{2}\left(1 - \sqrt{T_{min}}\right)
    \label{eqn:ke}
\end{equation}
The acoustic external coupling rate $\gamma_e$ can extracted from the linear $S_{11}$ reflection spectrum as,
\begin{equation}
    \gamma_e = \left(1 - S_{11}\right)\frac{\gamma}{2}
    \label{eqn:ye}
\end{equation}

We consider the acoustic resonator being driven by a single microwave frequency $\Omega_m$ such that,
\begin{equation}
b_{in} = B_{in}e^{-i \Omega_m t}
\label{eqn:bin}
\end{equation}
where $B_{in}$ is the amplitude of the input microwave field with power $P_{in} = \hbar\Omega_m |B_{in}|^2$ \cite{shao2019microwave}. In the weak optical mode limit, the optical backaction term $\left(ig_{0}a^{\dagger}a\right)$ can be neglected. Using value of $b_{in}$ from (\ref{eqn:bin}) in equation (\ref{eqn:HLb}) and using the Fourier transform, we get,
\begin{equation}
    B(\Omega) = \frac{\sqrt{\gamma_e}B_{in}2\pi\delta(\Omega + \Omega_m)}{i(\Omega+\Omega_m)+\frac{\gamma}{2}}
    \label{eqn:Bw}
\end{equation}
Now taking the inverse Fourier transform of (\ref{eqn:Bw}) gives us the solution to the equation (\ref{eqn:HLb}) as,
\begin{equation}
    b(t) = Be^{-i\Omega_m t} \: ; \:B = \frac{\sqrt{\gamma_e}B_{in}}{\gamma/2}
    \label{eqn:bt}
\end{equation}

\subsection{Optical sidebands}
The optical amplitude $a$ can be decomposed into a series of sidebands, such that,
\begin{equation}
    a = A_v \sum_{v}A_v e^{-iv\Omega_m t} \,,
    \label{eqn:Av}
\end{equation}
where $A_v$ is the amplitude of the optical sideband of order $v$. Under the limits of weak microwave input power and small optical power inside the cavity, we can ignore second and higher order optical sidebands, i.e. $v = 0, \pm 1$, which gives us,
\begin{equation}
    a = A_{0} + A_{-}e^{i\Omega_m t} + A_{+}e^{-i\Omega_m t}
    \label{eqn:sidebands}
\end{equation}
We can now solve for the amplitudes of the sidebands by using the value of $b$ from (\ref{eqn:bt}) in equation (\ref{eqn:HLa}). Applying Fourier transform to (\ref{eqn:HLa}) yields the following equation,
\begin{eqnarray}
    i\omega a(\omega) =&&\; \left(i\Delta - \frac{\kappa}{2}\right) a(\omega) - ig_{0}B \Big(a(\omega - \Omega_m) \nonumber \\
    +&&\; a(\omega + \Omega_m) \Big) + 2\pi\sqrt{\kappa_e}a_{in}\delta(\omega) 
    \label{eqn:Aw}
\end{eqnarray}
which can be decomposed into the following set of equations using (\ref{eqn:sidebands}) as,
\begin{eqnarray}
    0 = \left(i\Delta - \frac{\kappa}{2}\right)A_{0} -&&\; ig_{0}B\left(A_{-}+A_{+}\right) + \sqrt{\kappa_e}a_{in} \qquad \quad \\
    i\Omega_m A_{-} =&&\; \left(i\Delta - \frac{\kappa}{2}\right)A_{-} - ig_{0}BA_{0} \\
    -i\Omega_m A_{+} =&&\; \left(i\Delta - \frac{\kappa}{2}\right)A_{+} - ig_{0}BA_{0} 
\end{eqnarray}
Solving the above equations, we get the sideband amplitudes as,
\begin{eqnarray}
    A_{-} =&&\; \frac{ig_{0}BA_{0}}{i\left(\Delta - \Omega_m\right) - \kappa/2} \label{eqn:Astokes}\\
    A_{+} =&&\; \frac{ig_{0}BA_{0}}{i\left(\Delta + \Omega_m\right) - \kappa/2} \label{eqn:Aantistokes}\\
    A_{0} \simeq&&\; \frac{-\sqrt{\kappa_e} a_{in}}{i\Delta - \kappa/2} \label{eqn:A0}
\end{eqnarray}
In this work, $\Omega_m$ is comparable to $\kappa$ but $\kappa > \Omega_m$ for all acoustic frequencies. The maximum $S_{21}$ value is obtained when laser is parked at slope of the optical resonance transmission such that $\Delta = \pm\kappa/2$, as can be seen in Fig. \ref{fig:AOmod}. In our case, the pump laser is blue detuned from the optical resonance, i.e. $\Delta = \kappa/2$. Using this value of $\Delta$ in the equations (\ref{eqn:Astokes}) and (\ref{eqn:Aantistokes}), we get
\begin{eqnarray}
    A_{-} =&&\; \left[\frac{ig_{0}B}{i\left(\frac{\kappa}{2} - \Omega_m\right) - \frac{\kappa}{2}} \right] \left[\frac{-\sqrt{\kappa_e}a_{in}}{i\frac{\kappa}{2} - \frac{\kappa}{2}}\right] \nonumber \\
    =&&\; \frac{-ig_{0}B\sqrt{\kappa_e}a_{in}}{\frac{\kappa}{2}\left[\Omega_m + i\left(\Omega_m - \kappa\right)\right]} \label{eqn:stokes_amplitude}\\\nonumber\\
    A_{+} =&&\; \left[\frac{ig_{0}B}{i\left(\frac{\kappa}{2} + \Omega_m\right) - \frac{\kappa}{2}} \right] \left[\frac{-\sqrt{\kappa_e}a_{in}}{i\frac{\kappa}{2} - \frac{\kappa}{2}}\right] \nonumber \\
    =&&\; \frac{ig_{0}B\sqrt{\kappa_e}a_{in}}{\frac{\kappa}{2}\left[\Omega_m + i\left(\Omega_m + \kappa\right)\right]} \label{eqn:antistokes_amplitude}
\end{eqnarray}

\subsection{Extracting $g_0$ from the $S_{21}$ measurement}
According to the input-output theory of open quantum systems, the transmitted optical power ($T$) of an evanescently coupled unidirectional waveguide-resonator system is given by \cite{haus1984waves,aspelmeyer2014cavity},
\begin{equation}
    T = \hbar\omega_L|a_{in} - \sqrt{\kappa_e}a|^2
    \label{eqn:T}
\end{equation}
The output microwave voltage $V$ generated at the photodetector from the beating between the transmitted pump laser and the optical sidebands is given by $R_{PD}T$ (terms oscillating with $\Omega_m$), where $R_{PD}$ is the conversion gain of the photodetector (defined in V/W). Following some analysis, the expression for voltage is reduced to the following equation, as,
\begin{eqnarray}
   V \;&&= R_{PD}\hbar\omega_L\sqrt{\kappa_e} \:\Big[ \!\left(a_{in}^*A_{-}+a_{in}A_{+}^*\right)e^{i\Omega_m t} + \ldots \nonumber \\
    &&\ldots + \left(a_{in}A_{-}^*+a_{in}^*A_{+}\right)e^{-i\Omega_m t}\Big]
    \label{eqn:voltage}
\end{eqnarray}
The output of the photodetector is received at port 2 of the VNA where it is mixed with a local oscillator at $\Omega_m$ and then passed through a low-pass filter of bandwidth \SI{10}{\kilo\hertz}.
Now if we consider,
\begin{eqnarray}
   M \;&&= \left(a_{in}^*A_{-}+a_{in}A_{+}^*\right) \qquad \qquad \label{eqn:M}\\
   \mathrm{and,} \qquad  M^* \;&&= \left(a_{in}A_{-}^*+a_{in}^*A_{+}\right)
\end{eqnarray}

The voltage registered at port 2 of the VNA is then given by,
\begin{eqnarray}
|V| = R_{PD}\hbar\omega_L\sqrt{\kappa_e} |M|
\label{eqn:VM}
\end{eqnarray}

Using the values of $A_{-}$ and $A_{+}$ from equations (\ref{eqn:stokes_amplitude}) and (\ref{eqn:antistokes_amplitude}), respectively, in equation (\ref{eqn:M}), we obtain,
\begin{eqnarray}
    M \;&&= \frac{-i4g_{0}B\sqrt{\kappa_e}a_{in}^2}{\kappa} \left[\frac{\Omega_m - i\kappa}{\left(2\Omega_m^3 - \kappa^2\right) - i2\Omega_m\kappa}\right] \qquad \nonumber \\
    \;&&= \frac{4g_{0}B\sqrt{\kappa_e}a_{in}^2}{\kappa} \left[\frac{\kappa^3 - i \left(2\Omega_m^3 - \Omega_m\kappa^2\right)}{4\Omega_m^4 + \kappa^4}\right] \qquad \\
    \Rightarrow \ |M| \;&&= \frac{4g_{0}B\sqrt{\kappa_e}a_{in}^2}{\kappa}\sqrt{\frac{\Omega_m^2 + \kappa^2}{4\Omega_m^4 + \kappa^4}} \label{eqn:modM}
\end{eqnarray}

Plugging this value of $|M|$ in equation (\ref{eqn:VM}), we obtain the amplitude of the microwave voltage registered in $S_{21}$ signal as,
\begin{eqnarray}
    |V| = \frac{4g_{0}B\kappa_eR_{PD}P_{opt}}{\kappa}\sqrt{\frac{\Omega_m^2 + \kappa^2}{4\Omega_m^4 + \kappa^4}}
\end{eqnarray}
where, $P_{opt} = \hbar\omega_L a_{in}^2$, is the input optical power going into the cavity. The microwave power generated at the port 2 of the VNA is then given by,
\begin{eqnarray}
P_{out} \;&&= \frac{V^2}{2R_{load}} \\ \nonumber \\
\;&&= \frac{8g_{0}^2B^2\kappa_e^2R_{PD}^2P_{rec}^2}{\kappa^2 R_{load}} \left[\frac{\Omega_m^2 + \kappa^2}{4\Omega_m^4 + \kappa^4}\right]
\end{eqnarray}
where, $R_{load}$ (\SI{50}{\ohm}) is the impedance of the VNA and $P_{opt}$ is replaced by the optical power received at the photodetector ($P_{rec}$), accounting for the insertion losses of the device from the output of the cavity to the input of the photodetector. Finally, the acousto-optic transmission spectrum $S_{21}$ is defined as,
\begin{eqnarray}
S_{21} \;&&= P_{out}/P_{in} \nonumber \\
\;&&= \frac{32g_{0}^2\gamma_e\kappa_e^2R_{PD}^2P_{rec}^2}{\hbar\Omega_m\gamma^2\kappa^2R_{load}} \left[\frac{\Omega_m^2 + \kappa^2}{4\Omega_m^4 + \kappa^4}\right]
\label{eqn:S21}
\end{eqnarray}
using the value of $B$ from equation (\ref{eqn:bt}) and $P_{in} = \hbar\Omega_m B_{in}^2$. The above equation can be re-arranged to extract the value of $g_0$ from the $S_{21}$ transmission spectrum. The extracted $g_0$ has units of angular frequency (rad.Hz) but it is divided by $2\pi$ to get $g_0$ in Hz. The equation (\ref{eqn:S21}) is valid for both sideband-resolved and sideband-unresolved regimes when the pump laser is detuned by $\pm\kappa/2$ from the optical resonance frequency. In the case of sideband resolved regime, where $\Omega_m \gg \kappa$, we can write equation (\ref{eqn:S21}) as,
\begin{eqnarray}
S_{21} \simeq \frac{8g_{0}^2\gamma_e\kappa_e^2R_{PD}^2P_{rec}^2}{\hbar\Omega_m^3\gamma^2\kappa^2R_{load}}
\label{eqn:S21_resolved}
\end{eqnarray}
The equation (\ref{eqn:S21_resolved}) is also valid for for the case when the laser detuning $\Delta = \pm \Omega_m$.

\subsection{Photon number conversion efficiency ($\eta$)}
The intracavity photon number ($N_{\text{cav}}$) or the average number of photons circulating inside the cavity for the input optical power $P_{opt}$ is given by \cite{aspelmeyer2014cavity},
\begin{equation}
    N_{\text{cav}} = \left<a^{\dagger}a\right> = \frac{\kappa_e}{\Delta^2+(\kappa/2)^2} \: \frac{P_{opt}}{\hbar \omega_L}
    \label{eqn:Ncav1}
\end{equation}
In this work, $\Delta = \kappa/2$, which gives us,
\begin{equation}
    N_{\text{cav}} = \frac{2\kappa_eP_{opt}}{\kappa^2\hbar\omega_L}
    \label{eqn:Ncav}
\end{equation}
where, $P_{opt} = -4$ dBm in our case owning to 7 dBm of loss from the grating coupler and 4 dBm of loss between the fiber array and air for 7 dBm of input power from the laser.

The photon number conversion efficiency ($\eta$) is defined as the ratio of the number of optical sideband photons, which are coupling out of the cavity, to that of the number of input microwave photons. It can be expressed as,
\begin{eqnarray}
\eta = \frac{\left|\sqrt{\kappa_e}A_{-}\right|^2}{\left|B_{in}\right|^2} + \frac{\left|\sqrt{\kappa_e}A_{+}\right|^2}{\left|B_{in}\right|^2} \label{eqn:eta}
\end{eqnarray}
Using equations (\ref{eqn:bt}), (\ref{eqn:stokes_amplitude}) and (\ref{eqn:antistokes_amplitude}), the above equation can be written as,
\begin{eqnarray}
\eta =&&\; \frac{16g_0^2\gamma_e\kappa_e^2P_{opt}}{\kappa^2\gamma^2\hbar\omega_L}\left[\frac{1}{2\Omega_m^2+\kappa^2-2\Omega_m\kappa}\right] + \nonumber \\
&&\; \frac{16g_0^2\gamma_e\kappa_e^2P_{opt}}{\kappa^2\gamma^2\hbar\omega_L}\left[\frac{1}{2\Omega_m^2+\kappa^2+2\Omega_m\kappa}\right] \\ \nonumber \\
=&&\; \frac{4g_0^2}{\gamma\kappa}\:\frac{2\kappa_eP_{opt}}{\kappa^2\hbar\omega_L}\:\frac{2\kappa_e\kappa\gamma_e}{\gamma}\left[\frac{1}{2\Omega_m^2+\kappa^2-2\Omega_m\kappa}\right] + \nonumber \\
&&\; \frac{4g_0^2}{\gamma\kappa}\:\frac{2\kappa_eP_{opt}}{\kappa^2\hbar\omega_L}\:\frac{2\kappa_e\kappa\gamma_e}{\gamma}\left[\frac{1}{2\Omega_m^2+\kappa^2+2\Omega_m\kappa}\right] \\ \nonumber \\
=&&\; C_{om}\:N_{\text{cav}}\:\frac{2\kappa_e\kappa\gamma_e}{\gamma}\left[\frac{1}{2\Omega_m^2+\kappa^2-2\Omega_m\kappa}\right] + \nonumber \\
&&\; C_{om}\:N_{\text{cav}}\:\frac{2\kappa_e\kappa\gamma_e}{\gamma}\left[\frac{1}{2\Omega_m^2+\kappa^2+2\Omega_m\kappa}\right]
\end{eqnarray}
where, $C_{om}$ is the single-photon optomechanical cooperativity and $N_{\text{cav}}$ is the intracavity photon number. The experimental results for our devices have been tabulated in Table \ref{tab1}.

\section{Fabrication process} \label{sec:fab_process}

The device is fabricated on a \SI{220}{\nano\meter} thick Gallium Arsenide (GaAs) layer \cite{jiang2020suspended}, which is suspended in air by removing the underlying \SI{1}{\micro\meter} thick layer of Aluminium Gallium Arsenide (Al$_{0.6}$Ga$_{0.4}$As). The light is coupled onto the chip using integrated suspended grating couplers, which are partially etched to a depth of \SI{70}{\nano\meter} and have a radius of curvature of \SI{25}{\micro\meter}. The coupled light traverses inside a suspended rib waveguide, defined by etching the \SI{220}{\nano\meter} GaAs layer to a depth of \SI{120}{\nano\meter}. The waveguide has a width of \SI{0.54}{\micro\meter} with a total rib width of \SI{6}{\micro\meter}, making sure that the optical mode is tightly confined in the GaAs rib core. The ribs are tethered to the unsuspended GaAs layer using pillars which act like links on a suspension bridge. The waveguide is coupled to a microring resonator (gap \SI{350}{\nano\meter}), which is modified from a traditional circular ring resonator in order to maximise the coupling between the optical and the acoustic modes. 

We begin our device fabrication by spin coating PMMA resist on a cleaned sample and patterning the PMMA resist to define the metal electrodes for IDTs and contact pads. The electrodes are patterned such that they are aligned parallel to $\langle110\rangle$ set of directions. This is immediately followed by deposition of \SI{45}{\nano\meter} of gold layer with \SI{5}{\nano\meter} of chromium adhesion layer using thermal evaporation. PMMA is then lifted off the sample using acetone leaving only the metal layers behind. After a brief inspection of the IDTs, we spin coat the sample with hydrogen silsesquioxane (HSQ), which is a negative tone resist for electron-beam lithography. In the next step, we pattern the HSQ to define the waveguide, resonator and the grating couplers but we also use HSQ to cover the metal region to protect against dry etching. The patterned resist is developed using tetramethylammonium hydroxide (TMAH) and the GaAs layer is etched to a depth of \SI{70}{\nano\meter} using Ar/BCl$_3$ chemistry in an inductively coupled plasma- reactive ion etcher (ICP-RIE). 

By keeping the current HSQ layer intact, we spin coat another layer of HSQ and pattern with electron-beam lithography to cover the grating couplers in order to protect them from getting further etched. Using the same chemistry, we etch the GaAs layer another \SI{50}{\nano\meter} defining our waveguide ribs, which are now \SI{100}{\nano\meter} thin. We repeat the spin coating, lithography and etching steps one more time to etch the GaAs layer \SI{100}{\nano\meter} $+$ \SI{20}{\nano\meter} (over-etch) and open slots for undercutting, while protecting the waveguide ribs. All the etch steps are monitored with an ellipsometer to ensure the etch depths within $\pm$\SI{5}{\nano\meter} are achieved. Finally, the device is suspended by wet etching the underlying Al$_{0.6}$Ga$_{0.4}$As layer with (24$\%$) hydrofluoric acid (HF). To remove any by-products and smoothing the GaAs walls, the sample is cleaned in diluted hydrogen peroxide (H$_2$O$_2$ - 30$\%$) and potassium hydroxide (KOH:H$_2$O = 1g:4mL) solutions \cite{midolo2015soft, jiang2020suspended}. In practice, we observed that performing the full cleaning step as in our previous passive devices \cite{jiang2020suspended} was affecting the RF performance of the IDTs, with the $S_{11}$ dips diminishing significantly post-fabrication. On the other hand, not cleaning the chip post-undercut was leaving the GaAs surface with a lot of surface residue. A trade off was made to keep the cleaning times brief, 10 seconds and 1 minute, respectively, which was used for the devices reported in this work.

\section{Fabricated rib waveguides} \label{sec:width}
\begin{figure*}[htbp]
\includegraphics[scale=1.0]{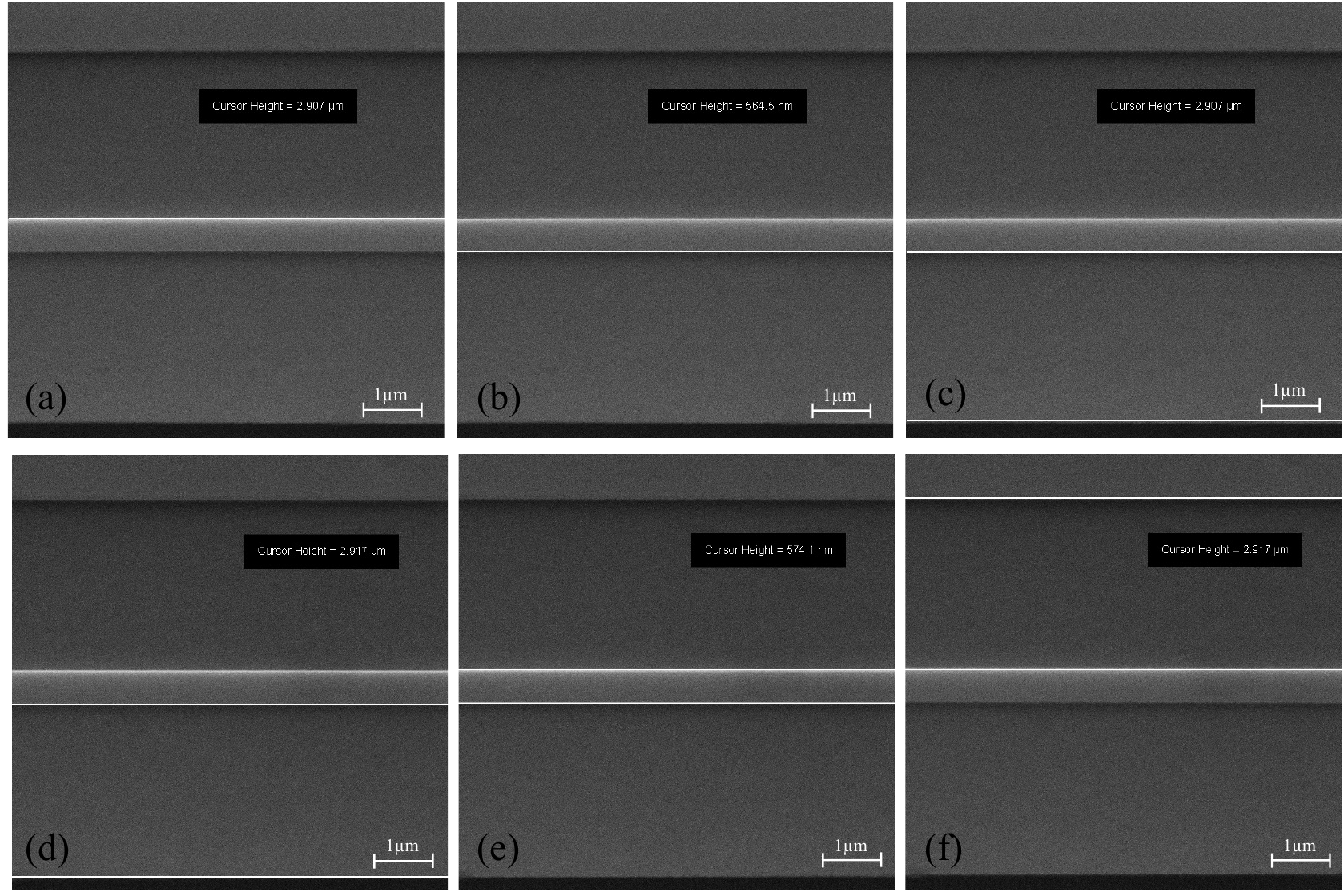}
\caption{SEM images of the rib waveguides actuated by the Lamb wave resonators. (a) - (c) Waveguide and rib widths measured for the device actuated by the IDT with a period of \SI{0.8}{\micro\meter}, whose measured spectra are depicted in Figures \ref{fig:g0_S21}(a) and \ref{fig:g0_S21_zoom}(a). (d) - (f) Waveguide and rib widths measured for the device actuated by the IDT with a period of \SI{1.0}{\micro\meter}, whose measured spectra are depicted in Figures \ref{fig:g0_S21}(b) and \ref{fig:g0_S21_zoom}(b).}
\label{fig:width}
\end{figure*}
The waveguide widths and the rib widths for the fabricated devices used in the measurements in Figs.\ref{fig:g0_S21} and Fig.\ref{fig:g0_S21_zoom} are estimated using SEM imaging. Figures \ref{fig:width}(a)-(c) depict the dimensions of the rib waveguide which has been actuated by an IDT of period \SI{0.8}{\micro\meter} and whose measurements and numerical calculations have been shown in Figures \ref{fig:g0_S21}(a) and \ref{fig:g0_S21_zoom}(a), while figures \ref{fig:width}(d)-(f) depict the dimensions of the rib waveguide which has been actuated by an IDT of period \SI{1.0}{\micro\meter} and whose measurements and numerical calculations have been shown in Figures \ref{fig:g0_S21}(b) and \ref{fig:g0_S21_zoom}(b). The waveguide dimensions have been used in numerical calculation of $g_0$ and have been shown by blue bars in figures \ref{fig:g0_S21} and \ref{fig:g0_S21_zoom}. The widths measured using SEM imaging have an error of $\pm$\SI{25}{\nano\meter} due to the manual cursor placement.

\section{AO-$S_{21}$ spectra for the mechanical supermode} \label{sec:3devices}
\begin{figure}[htbp]
\includegraphics[scale=0.45]{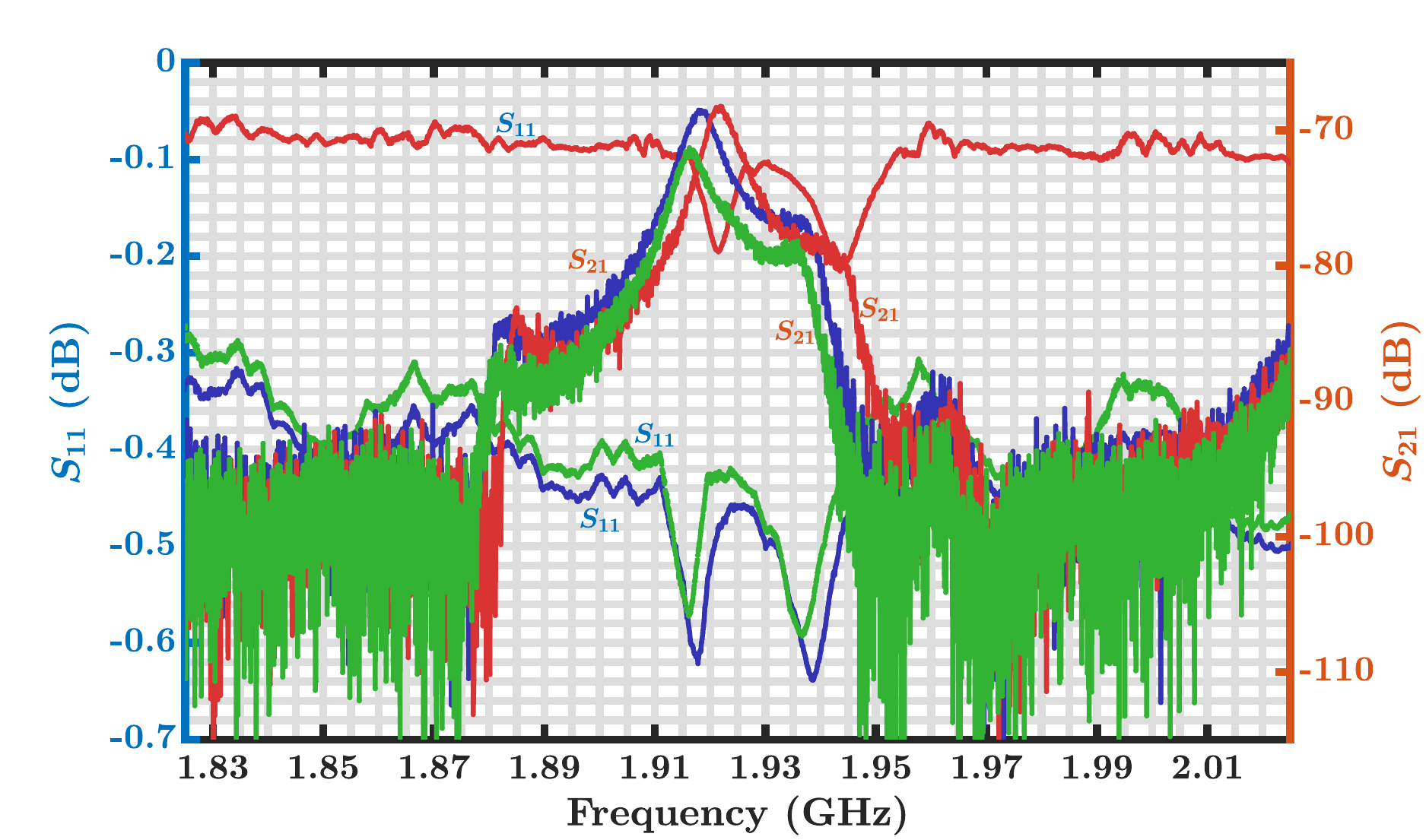}
\caption{Measured AO-$S_{21}$ spectra for three devices at $\approx$ \SI{1.92}{\giga\hertz}, each actuated by an IDT with a period of \SI{0.8}{\micro\meter}. The RF reflection ($S_{11}$) spectra for each are also shown in the same color for reference.}
\label{fig:S21_3_devices}
\end{figure}
Figure \ref{fig:S21_3_devices} shows the AO-$S_{21}$ spectra for three devices actuated by IDTs with a period of \SI{0.8}{\micro\meter}. In all three case, a doublet in the $S_{11}$ spectra has been observed and an enhancement in transduction efficiency ($\eta_{peak}$) has been obtained for the mechanical mode at $\approx$ \SI{1.92}{\giga\hertz}. This shows that the supermode formation is reasonably robust to fabrication errors. As long as the modes are lined up within the IDT transduction bandwidth, $\eta_{peak}$ is significantly enhanced. The sensitivity of supermode formation can be estimated by looking at the three curves, which correspond to nominally identical devices. The green curve in Fig.\ref{fig:S21_3_devices} shows an $\eta_{peak}$ reduced by $\approx$ 3 dB for a shift of $\approx$ 5.4 MHz in frequency.

\section{Factors influencing optical performance} \label{sec:loss}
\begin{figure}[!htbp]
\includegraphics[scale=0.45]{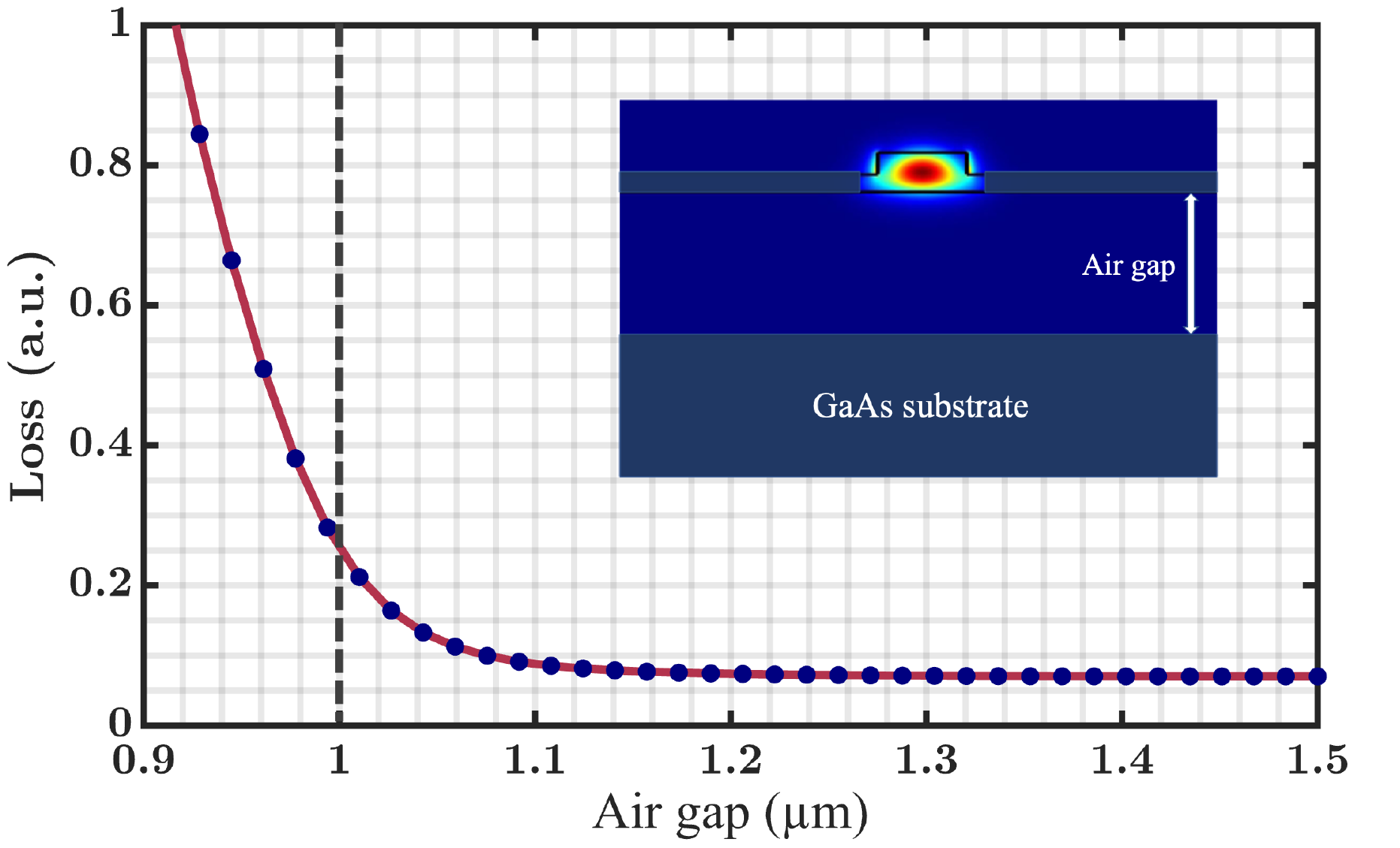}
\caption{\label{fig:loss} Optical loss obtained from mode solving in Ansys Lumerical for varied thickness of bottom air cladding.}
\end{figure}

To better understand the factors affecting the optical loss observed in our devices, in particular the lower than expected $Q_o$, we simulated our rib waveguide structure (shown in the inset of Fig.\ref{fig:loss}), using Ansys Lumerical's Mode solutions. The structure has the waveguide width of \SI{540}{\nano\meter} and the rib width of \SI{6}{\micro\meter} with the bottom cladding (in our case, air) varied between \SI{0.92}{\micro\meter} - \SI{1.5}{\micro\meter} thickness. As we can see from Fig.\ref{fig:loss}, the relative loss (normalized to a gap of 920 nm) increases exponentially as the gap between the waveguide and the substrate goes down, due to the leakage of the propagating optical mode into the substrate. Our devices were fabricated on a \SI{220}{\nano\meter} GaAs layer, which was suspended by removing the underlying \SI{1}{\micro\meter} of Aluminium Gallium Arsenide (Al$_{0.6}$Ga$_{0.4}$As) layer. As we can see from Fig.\ref{fig:loss}), losses increase by roughly $15\%$ as the air gap decreases from \SI{1.1}{\micro\meter} to \SI{1}{\micro\meter}. Since our devices are suspended in air and GaAs has relatively low tensile stress, the effective air gap between the waveguide and the substrate is possibly lower than \SI{1}{\micro\meter}, something we had observed in previous experiments as reported in \cite{jiang2020suspended}. As the loss increases exponentially with decreasing air gap, this leads to higher optical loss and lower $Q_o$ for the devices. By increasing the AlGaAs thickness to \SI{1.5}{\micro\meter}, we can significantly reduce the optical losses while also allowing for sufficient tolerance post fabrication.

{\color{black} \section{3D FEM simulations of the mechanical supermode} \label{sec:3Dsim}
\begin{figure}[H]
\includegraphics[scale=0.56]{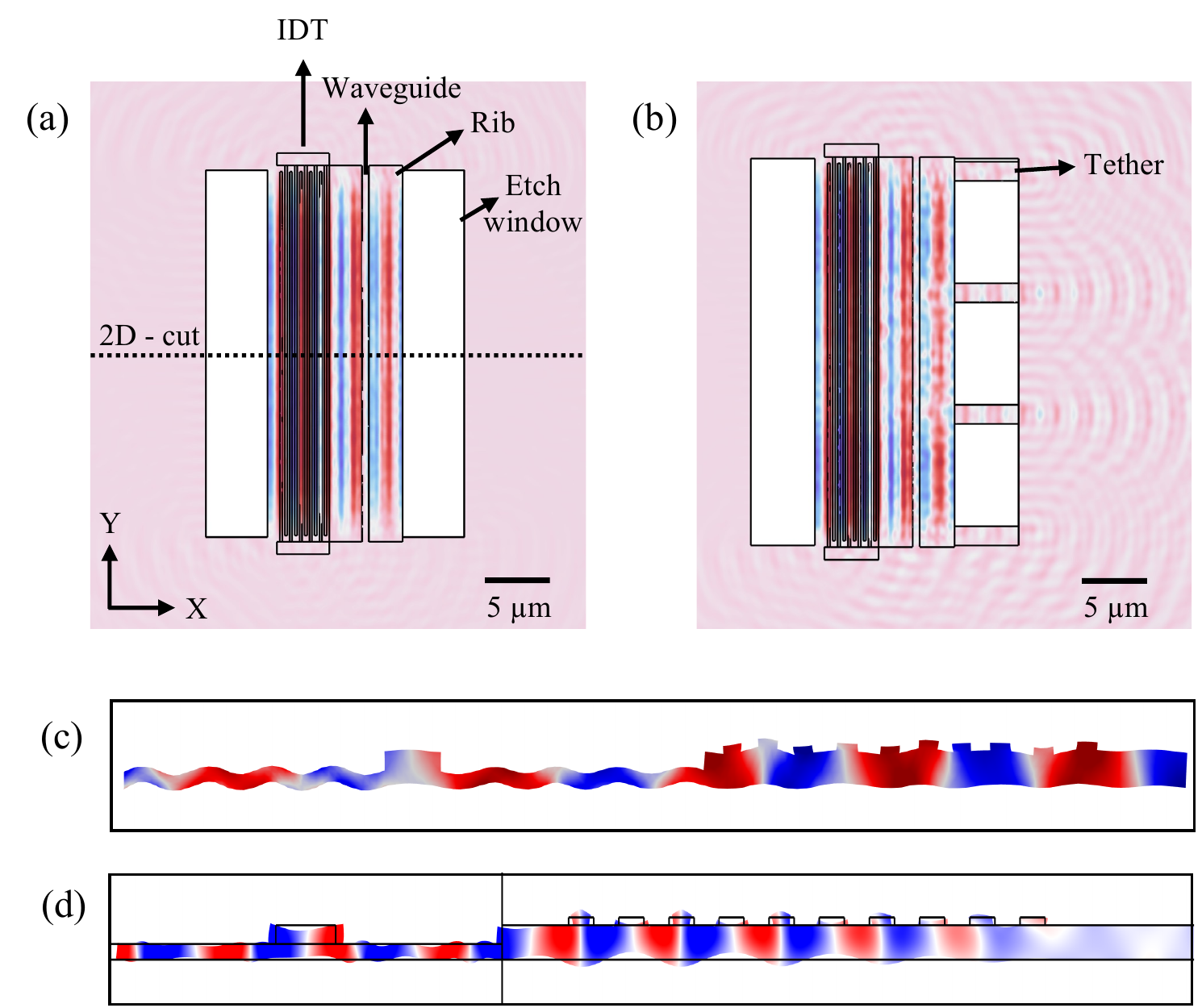}
\caption{\kcb{3D simulations of the supermode formation between the contour mode resonator and the shear horizontal (SH) mode of the rib waveguide. (a) X displacement of the SH breathing mode at \SI{1.97}{\giga\hertz} generated by the contour mode resonator (IDT), which is coupled to a rib waveguide. (b) X displacement of the SH mechanical mode at
\SI{1.97}{\giga\hertz} when the rib waveguide is held by the tethers between the etch window. A scaled version of the device (interaction length = \SI{30}{\micro\meter}) shown in Fig.2 is used to fit the simulation in memory. The slight distortion to the mode shape at the tether locations and the acoustic radiation through the tethers can be clearly seen. (c) 2D cross-section of the mode displacement of the simulated mechanical mode at \SI{1.97}{\giga\hertz} shown in (a). It can be seen that the mode profile of this 2D cut is very similar to the mechanical mode profile of Fig.\ref{fig:eigenmodes}(g), shown in (d) for reference. At the supermode frequency, the contour mode resonator clearly excites the SH mode of the rib waveguide.}}\label{fig:3Dsim} 
\end{figure}

While the supermode formation and the acousto-optic overlap integrals were calculated with 2D simulations for computational tractability, it is an important question to understand the effect of the 3D shape of the resonator on the supermode formation and the role of the tethers. Fig.\ref{fig:3Dsim}(a) shows a 3D FEM simulation of the contour mode resonator, coupled to the rib waveguide excited at a frequency of \SI{1.97}{\giga\hertz} corresponding to supermode formation in this device. The $X$-displacement of the mode shape is shown in Fig.\ref{fig:3Dsim}(a). A scaled version of the actual fabricated device with interaction length of \SI{30}{\micro\meter}, and 5 finger pairs is chosen to fit the computation within memory. Fig.\ref{fig:3Dsim}(b) shows the same structure but with suspension tethers added. The spacing between the tethers and the tether widths in the simulation are half of that in the fabricated device. As can be seen from Fig.\ref{fig:3Dsim}(b), the overall modeshape at the supermode frequency is nominally identical to that in Fig.\ref{fig:3Dsim}(a), except that there is a distortion of the mode shape near the tethers and significant energy leakage, as can be seen by the acoustic radiation patterns at the tether ends.  

The correspondence between the 2D mechanical mode simulations used for determining the acousto-optic overlaps and the 3D mode shapes can be determined by taking a cross-section of the mode displacement in Fig\ref{fig:3Dsim}(a). Fig.\ref{fig:3Dsim}(c) shows the corresponding mode displacement. The supermode, as determined from a 2D eigenmode simulation, Fig.1(g), is reproduced in Fig.\ref{fig:3Dsim}(d). The correspondence between the two can be clearly seen, showing the 3D simulations also indicate the existence of supermode formation between the SH breathing mode of the rib waveguide and the Lamb wave mode of the contour mode resonator. Moving forward, optimizing the design and placement of the suspension tethers to minimize mode perturbation and leakage is critically important to achieve high overall transduction efficiency.}

\nocite{*}

\bibliography{References}

\end{document}